\documentclass[fleqn,usenatbib]{mnras}

\usepackage{newtxtext,newtxmath}

\usepackage[T1]{fontenc}

\DeclareRobustCommand{\VAN}[3]{#2}
\let\VANthebibliography\thebibliography
\def\thebibliography{\DeclareRobustCommand{\VAN}[3]{##3}\VANthebibliography}

\usepackage{graphicx}	
\usepackage{amsmath}	

\usepackage{orcidlink}

\title[MIGHTEE radio galaxy bias from CMB lensing]{Galaxy bias of radio AGN and star-forming galaxies from MIGHTEE and CMB lensing}

\author[J. Hamlett et al.]{Joel Hamlett$^{\orcidlink{0009-0002-9611-1970}}$,$^{1}$\thanks{E-mail: joel.hamlett@physics.ox.ac.uk}
Catherine L. Hale$^{\orcidlink{0000-0001-6279-4772}}$,$^{2}$
Matt J. Jarvis$^{\orcidlink{0000-0001-7039-9078}}$,$^{1,3}$
David Alonso$^{\orcidlink{0000-0002-4598-9719}}$,$^{1}$
S. Lyla Jung$^{\orcidlink{0000-0001-5512-3735}}$,$^{1}$
\newauthor
Natalia Stylianou$^{\orcidlink{0000-0002-6029-005X}}$$^{1}$ and
Rohan G. Varadaraj$^{\orcidlink{0009-0006-9953-6471}}$$^{1}$
\\
\\
$^{1}$Astrophysics, University of Oxford, Denys Wilkinson Building, Keble Road, Oxford, OX1 3RH, UK\\
$^{2}$School of Physics and Astronomy, Institute for Astronomy, University of Edinburgh, Royal Observatory, Blackford Hill, EH9 3HJ Edinburgh, UK\\
$^{3}$Department of Physics and Astronomy, University of the Western Cape, Robert Sobukwe Road, 7535 Bellville, Cape Town, South Africa\\
}

\date{Accepted XXX. Received YYY; in original form ZZZ}

\pubyear{\the\year{}}

\begin{document}
\label{firstpage}
\pagerange{\pageref{firstpage}--\pageref{lastpage}}
\maketitle

\begin{abstract}
Radio galaxies play a significant role in galaxy evolution through feedback mechanisms, making it important to characterize the environments in which they reside. We measure the cross-correlation between the positions of radio sources in the MIGHTEE survey and cosmic microwave background lensing measurements from the Atacama Cosmology Telescope over $20\ {\rm deg}^2$ to flux densities $S_{\rm 1.4\,GHz}>40\ {\rm \mu Jy}$, detecting the signal at $\sim4\sigma$ significance. By combining this with auto-correlation measurements of low- and high-redshift radio samples, we constrain the redshift evolution of the galaxy bias parameter, $b(z)$. Fitting different models for the redshift evolution of the bias, we find that both a model inversely proportional to the linear growth factor, $b(z)=(1.63^{+0.19}_{-0.22})/D(z)$, and a model where bias grows linearly with redshift, $b(z)=(0.49^{+0.44}_{-0.32})+(1.62^{+0.37}_{-0.51})z$, provide comparably good fits to the data, with the current measurements unable to distinguish between them. Using a radio luminosity function, we estimate the fraction of sources that are active galactic nuclei (AGNs) or star-forming galaxies (SFGs) as a function of redshift, and model the individual contributions of AGNs and SFGs to the total bias. This approach allows us to measure the bias of AGNs and SFGs without individual source classifications. Though no single model is favoured significantly, the best-fitting model gives $b_{\rm AGN}(z) = (2.60^{+0.65}_{-0.91})/D(z)$ compared to $b_{\rm SFG}(z) = (0.52^{+0.47}_{-0.34})/D(z)$. Incorporating SFG bias measurements from previous work as a prior tightens these constraints to $b_{\rm AGN}(z) = (1.96^{+0.37}_{-0.40})/D(z)$ and $b_{\rm SFG}(z) = (0.90^{+0.06}_{-0.06})/D(z)$.
\end{abstract}




\section{Introduction}
\label{sec:introduction}

Radio continuum surveys provide a powerful probe of the large-scale structure of the Universe, tracing galaxy populations to high redshift with minimal dust attenuation. The radio sky below frequencies of $\sim10\ {\rm GHz}$ is dominated by synchrotron emission \citep{condon1992} originating from two classes of objects: active galactic nuclei (AGNs) and star-forming galaxies (SFGs). 
Radio-AGNs can be broadly divided into radio-loud and radio-quiet populations \citep{kellermann, wilson1995}. The radio emission from radio-loud AGNs is associated with powerful relativistic jets launched by accretion onto the supermassive black hole at the centre of a galaxy \citep[e.g.,][]{best2012, heckman2014}, whereas the origin of the weaker radio emission in radio-quiet AGNs is less certain \citep{white2015, white2017, panessa2019}. Through feedback mechanisms, AGNs are thought to play an important role in regulating both their host galaxy evolution and the surrounding environment \citep{croton2006, bower2006, fabian2012, hardcastle2020}. Radio-AGNs are typically hosted by the most massive galaxies \citep[e.g.,][]{best2005, smolcic2009, gurkan2014, sabater2019, capetti2022, best2023, kondapally2025}, and are preferentially located in overdense environments such as galaxy groups and clusters \citep[e.g.,][]{hickox2009, mandelbaum2009, lindsay2014, magliocchetti2017, hale2018, uchiyama2022}. In contrast, the abundance of SFGs decreases with local density \citep{best2004, sabater2013}, and they show no environmental differences relative to their radio-quiet counterparts \citep{bardelli2010, magliocchetti2018}. Previous studies have consistently found that radio-AGNs reside in significantly more strongly clustered environments than radio-selected SFGs \citep[][see the review by \citealt{magliocchetti2022}]{magliocchetti2017, hale2018, chakraborty2020, mazumder2022}.


Radio-loud AGNs can be further classified according to their accretion efficiency into high-excitation radio galaxies (HERGs), which have high-excitation lines visible in their optical spectra \citep{hine1979}, and low-excitation radio galaxies \citep[LERGs; see the reviews by][]{heckman2014, hardcastle2020}. HERGs are thought to be powered by radiatively efficient accretion processes fuelled by cold gas reservoirs \citep{hardcastle2007}. They are typically hosted by less massive galaxies with higher star-formation rates \citep{best2012, hardcastle2013, gurkan2015}. In contrast, the majority of LERGs are assumed to be powered by radiatively inefficient accretion, likely sustained by advection dominated flows from their host galaxy haloes and surrounding environment \citep{narayan1995, kording2006}. While some studies have found significant overlap in the accretion efficiencies of these populations at low flux densities \citep{whittam2018, whittam2022}, more recent work has recovered a clear difference when considering only sources with high-confidence spectroscopic classifications \citep{arnaudova2025}. LERGs are generally hosted by massive red galaxies with older stellar populations \citep{smolcic2009, herbert2010, ching2017b, williams2018} and are more likely to be found at the centres of galaxy groups or clusters \citep{best2007}. Consequently, LERGs tend to reside in richer environments and are more strongly clustered than HERGs \citep{hardcastle2004, ineson2015, ching2017, hale2018, lopez2026}.

Galaxies are biased tracers of the underlying matter distribution since they form within dark matter haloes and therefore preferentially sample denser regions of the matter field \citep{white1978}. The relationship between the galaxy and matter overdensities can be described by the galaxy bias parameter, $\Delta_{\rm g} = b_{\rm g} \Delta_{\rm m}$ \citep{peebles1980, kaiser1984}. Studying the bias allows us to better understand the environments galaxies inhabit and the evolution of the galaxy-halo connection. Measurements of the bias of different galaxy populations can be used to compare their environments and, when combined, support multi-tracer techniques that mitigate cosmic variance at large scales \citep{raccanelli2012, ferrmacho2014, gomes2020}. 

A common method to estimate the bias of radio galaxies is to measure their auto-correlation signal \citep{magliocchetti2017, hale2018, chakraborty2020, mazumder2022, hale2024} or the cross-correlation of radio sources with another galaxy population \citep{lindsay2014, hale2025b, hamlett2026}. An alternative and complementary approach is to cross-correlate the distribution of radio sources with maps of gravitational lensing of the cosmic microwave background (CMB). CMB photons originate from the last-scattering surface at $z\approx1100$, and their trajectories are gravitationally deflected by the large-scale matter distribution along the line of sight \citep{bartelmann2001}. The resulting CMB lensing signal is most sensitive to structures at redshift $z\sim2$ which makes radio surveys particularly well suited for cross-correlation studies since radio emission is largely unaffected by dust attenuation and can probe galaxy populations out to high redshift. Since CMB lensing traces the underlying matter distribution directly, it alleviates the degeneracy between the bias and the amplitude of the matter power spectrum that arises in galaxy-galaxy cross-correlations.

The majority of previous cross-correlation studies with CMB lensing maps have used wide-area radio surveys covering $\sim10^2-10^4\ {\rm deg^2}$ \citep{smith2007, allison2015, alonso2021, piccirilli2023, nakoneczny2024, tanidis2025}. In this work, we instead exploit the depth of the MeerKAT International GHz Tiered Extragalactic Exploration (MIGHTEE) Survey \citep{jarvis2016}. We use the first data release \citep[DR1;][]{hale2025} which covers three fields with a total area of $\sim20\ {\rm deg^2}$ at an average effective frequency of $\sim1.3\ {\rm GHz}$. Although the survey area is smaller than those used in previous studies, the greater depth of MIGHTEE allows access to fainter radio populations and partially compensates for the reduced sky coverage in terms of effective volume.

Accurate modelling of the auto- and cross-correlation signals requires knowledge of the redshift distribution of the radio sources, which cannot be obtained from radio observations alone. Previous clustering studies have used a variety of approaches to estimate these distributions. For example, \cite{nusser2015}, who measured the auto-correlation of radio sources from the NRAO VLA Sky Survey \citep[NVSS;][]{condon1998}, relied on redshifts that were available for a small subsample of objects (131 out of $\sim6\times10^5$) from the Hercules \citep{waddington2001} and CENSORS \citep{best2003, rigby2011} surveys. \cite{nakoneczny2024}, who cross-correlated sources from the second data release of the LOFAR Two-metre Sky Survey \citep[LoTSS DR2;][]{shimwell2022} with CMB lensing, instead inferred the redshift distribution of the wide-area sample using deeper multi-wavelength observations over smaller fields, while \cite{tanidis2025} adopted the redshift distribution from simulated radio catalogues when cross-correlating sources from the Evolutionary Map of the Universe survey \citep[EMU;][]{norris2021}. A different approach was taken by \cite{alonso2021}, who used measurements of the auto-correlation of LoTSS DR1 sources \citep{vanhaarlem2013, shimwell2019} and their cross-correlation with CMB lensing to break degeneracies and constrain the high-redshift tail of the redshift distribution from the correlation signals themselves. The MIGHTEE fields benefit from rich, overlapping multi-wavelength ancillary data over a large fraction of the area used in this work. The radio catalogues have been cross-matched with their near-infrared host galaxies \citep{hale2026}, providing redshift estimates for $\sim90$ per cent of the radio sources in the overlap region.

In this work, we measure the auto-correlations of low- and high-redshift radio samples, and the cross-correlation between the radio source distribution and the CMB lensing convergence (defined in Section~\ref{sec:theory}) maps from the Atacama Cosmology Telescope \citep[ACT;][]{fowler2007, swetz2011} sixth data release \citep[DR6;][]{qu2024, madhavacheril2024, maccrann2024, naess2025}. From these measurements, we constrain the effective galaxy bias of the radio population. Previous CMB lensing cross-correlation studies have either measured the bias of the mixed radio population \citep[i.e., combined SFGs and AGNs;][]{alonso2021, piccirilli2023, nakoneczny2024, tanidis2025}, or of an AGN-dominated sample selected using a high flux density threshold \citep[$\gtrsim1\ {\rm mJy}$;][]{allison2015}. Identifying AGNs and SFGs individually requires additional information, such as redshifts to determine radio luminosities \citep[e.g.,][]{magliocchetti2014b, hamlett2026}, or multi-wavelength diagnostics \citep[e.g.,][]{smolcic2017, whittam2022, best2023}, which reduces the available sample sizes.

Instead, we model the radio population using $1.4\ {\rm GHz}$ radio luminosity functions to estimate the fraction of sources that are AGNs or SFGs as a function of redshift. This allows the measured effective bias to be decomposed into its AGN and SFG contributions by modelling the bias of each population separately. In this way, we can constrain the clustering properties of AGNs and SFGs while retaining the full radio source sample down to flux densities of $40\ {\rm \mu Jy}$.

The paper is structured as follows: in Section~\ref{sec:theory} the theory behind the correlation signals is discussed. In Section~\ref{sec:data} we discuss the radio data, obtaining redshift distributions, constructing the low- and high-redshift radio samples, and the ACT lensing maps. Section~\ref{sec:methods} describes the methods used to obtain the correlation signals. In Section~\ref{sec:results} we present the measured correlation signals, perform robustness tests, and discuss their significance. The results are summarized in Section~\ref{sec:conclusions}. Throughout this paper, we assume a flat $\Lambda$CDM cosmology with the standard cosmological parameters from the \cite{planck2020}: $H_0=67.4\ {\rm km\ s^{-1}\ Mpc^{-1}}$, $\Omega_{\rm b}=0.0493$, $\Omega_{\rm c}=0.264$, $\sigma_8=0.811$ and $n_{\rm s}=0.965$. We also assume a constant spectral index for radio sources, $\alpha=-0.7$, with $S_{\nu} \propto \nu^\alpha$ for a source with flux density $S_\nu$ at frequency $\nu$.

\section{Theory}
\label{sec:theory}

\subsection{Projected tracers of large-scale structure}

We analyse the statistical relationship between two projected fields on the sky: the galaxy number overdensity, $\delta_{\rm g}$, and the CMB lensing convergence, $\kappa$. The galaxy overdensity is defined as the fractional fluctuations of the observed number of galaxies, $N_{\rm g}$, in a given sky direction, $\hat{\mathbf{n}}$, relative to the sky-averaged value, $\overline{N}_{\rm g}$:
\begin{equation}
\label{eqn:galaxy_overdensity_definition}
    \delta_{\rm g}(\hat{\mathbf{n}}) = \frac{N_{\rm g}(\hat{\mathbf{n}}) - \overline{N}_{\rm g}}{\overline{N}_{\rm g}}.
\end{equation}
Assuming a known redshift distribution, $p(z)$, normalized to unity when integrated over $z$, the observed overdensity can be written as the projection of the three-dimensional galaxy density constrast, $\Delta_{\rm g}$:
\begin{equation}
    \delta_{\rm g}(\hat{\mathbf{n}}) = \int {\rm d}z \, p(z) \Delta_{\rm g}(\chi(z)\hat{\mathbf{n}}, z),
\end{equation}
where $\chi(z)$ is the comoving radial distance. On the large scales probed in this work (comoving wavenumber $k \lesssim 0.15 \ {\rm Mpc}^{-1}$), we adopt a linear, scale-independent bias model, $b_{\rm g}(z)$, relating the galaxy overdensity and matter overdensity, $\Delta_{\rm m}$, at comoving position $\mathbf{x}$:
\begin{equation}
    \Delta_{\rm g}(\mathbf{x}, z) = b_{\rm g}(z) \Delta_{\rm m}(\mathbf{x}, z).
\end{equation}

The CMB lensing convergence quantifies the distortion of CMB photon trajectories induced by the gravitational potential of large-scale structure encountered along the line of sight between the last scattering surface and the observer. It is defined to be proportional to the divergence of the deflection angle field, $\boldsymbol{\alpha}$:
\begin{equation}
    \kappa(\hat{\mathbf{n}})=-\frac{1}{2} \boldsymbol{\nabla} \cdot \boldsymbol{\alpha}(\hat{\mathbf{n}}) = -\frac{1}{2} \nabla^2 \phi(\hat{\mathbf{n}}),
\end{equation}
where $\phi$ is the lensing potential. The lensing convergence can be expressed as a weighted line-of-sight projection of the matter overdensity field:
\begin{equation}
    \kappa(\hat{\mathbf{n}}) = f_\ell \int_0^{\chi_{\rm LSS}} {\rm d}\chi  \,  \frac{3 H_0^2 \Omega_{\rm m, 0}}{2a} \frac{(\chi_{\rm LSS}-\chi)\chi}{\chi_{\rm LSS}} \Delta_{\rm m}(\chi\hat{\mathbf{n}}, z),
\end{equation}
where $H_0$ is the Hubble constant, $\Omega_{\rm m, 0}$ the present-day matter density parameter, $a=1/(1+z)$ is the scale factor, and $\chi_{\rm LSS}$ the comoving distance to the last scattering surface ($z\approx1100$). The multipole, $\ell$, dependent factor, $f_\ell$, is given by
\begin{equation}
    f_\ell = 1 - \frac{1}{(2\ell+1)^2},
\end{equation}
and accounts for the relation between the convergence and the angular Laplacian of the lensing potential. It only affects the largest angular scales ($\ell \lesssim 10$).

\subsection{Angular power spectra of projected fields}

Both observables considered here, $\delta_{\rm g}$ and $\kappa$, can be written generically as projected fields on the sphere:
\begin{equation}
    u(\hat{\mathbf{n}}) = \int {\rm d}\chi \, W_u(\chi) \Delta_{\rm m}(\chi \hat{\mathbf{n}}, z(\chi)),
\end{equation}
where $W_u(\chi)$ denotes the corresponding radial kernel. Expanding these projected fields into spherical harmonics, 
\begin{equation}
    u(\hat{\mathbf{n}}) = \sum_{\ell,m} u_{\ell m} Y_{\ell m}(\hat{\mathbf{n}}),
\end{equation}
their statistical properties are fully characterized by the angular power spectra, $C_\ell^{uv}$:
\begin{equation}
    \left\langle u_{\ell m}v^\ast_{\ell' m'} \right\rangle = \delta_{\ell \ell'} \delta_{m m'} C_\ell^{uv}.
\end{equation}
Under the Limber approximation \citep{limber1954}, which is valid when the projection kernels vary slowly compared to the typical correlation length of the density field, as is the case for the kernels in this work, the angular power spectrum can be related to the three-dimensional matter power spectrum, $P_{\rm mm}(k, z)$, via
\begin{equation}
    C_\ell^{uv} = \int \frac{{\rm d}\chi}{\chi^2} \, W_u(\chi) W_v(\chi) P_{\rm mm} \left( k=\frac{\ell+1/2}{\chi}, z \right).
\end{equation}
The matter power spectrum is defined through
\begin{equation}
    \left\langle \Delta_{\rm m}(\mathbf{k}, z) \Delta_{\rm m}^\ast(\mathbf{k}', z) \right\rangle = (2\pi)^3 \delta^3(\mathbf{k}-\mathbf{k}') P_{\rm mm}(k, z),
\end{equation}
with the Fourier convention
\begin{equation}
    \Delta_{\rm m}(\mathbf{x}, z) = \int \frac{{\rm d}^3 k }{(2\pi)^3} e^{i \mathbf{k} \cdot \mathbf{x}} \Delta_{\rm m}(\mathbf{k}, z).
\end{equation}
For the two observables considered in this work, the relevant kernels are
\begin{equation}
\label{eqn:galaxy_kernel}
    W_{\rm g}(\chi) = \frac{H(z)}{c} p(z) b_{\rm g}(z),
\end{equation}
where $H(z)$ is the Hubble parameter and $c$ is the speed of light, and 
\begin{equation}
    W_{\kappa}(\chi) = f_\ell \frac{3 H_0^2 \Omega_{\rm m, 0}}{2a} \frac{(\chi_{\rm LSS}-\chi)\chi}{\chi_{\rm LSS}} \Theta(\chi-\chi_{\rm LSS}),
\end{equation}
with $\Theta$ the Heaviside step function.

Since $\delta_{\rm g}$ is defined from observed counts, $W_{\rm g}$ in principle also receives a contribution from magnification bias, which changes the number of sources above the flux limit by brightening and distorting sources. We measure the response of the number counts to the magnitude/flux limit and find $ {\rm d} \log_{10}{N} / {\rm d} (-2.5\log_{10}{S_{\rm lim}}) \approx 0.39$ which is close to $0.4$ at which magnification bias is zero \citep{moessner1998}, so we neglect this contribution in our modelling. 

All theoretical calculations in this work were performed using the Core Cosmology Library \citep[\textsc{CCL};][]{chisari2019}. To calculate the matter power spectrum, $P_{\rm mm}(k, z)$, we use the \textsc{HALOFIT} \citep{smith2003, takahashi2012} parametrization as implemented in the \textsc{CAMB} Boltzmann code \citep{lewis2000}.

\section{Data}
\label{sec:data}

\subsection{MIGHTEE DR1}
\label{ssec:mightee}

The radio sources in this work are from the first data release (DR1) of the MIGHTEE radio continuum survey \citep{hale2025}. The survey was conducted using the MeerKAT radio telescope \citep{jonas2016} and covers three widely-separated fields: $4.2 \ {\rm deg}^2$ of the Cosmic Evolution Survey field (COSMOS), $14.4 \ {\rm deg}^2$ of the XMM Large-Scale Structure field (XMM-LSS), and deeper imaging over $1.5 \ {\rm deg}^2$ of the Extended Chandra Deep Field South (CDFS). Both higher- ($\sim5''$) and lower-resolution ($\sim7-9''$) images are available, and we use the catalogues derived from the deeper lower-resolution images.

Source detection was performed using the source finding software \texttt{PyBDSF} \citep{mohan2015} using an iterative process to improve source recovery from the confusion limited images \citep[see Section 3 of][]{hale2025}. As a part of this process, \texttt{PyBDSF} generates local root mean square (r.m.s.) sensitivity maps, from which the lower-resolution images have median sensitivities of $3.5$, $3.2$ and $2.0\ {\rm \mu Jy \ beam^{-1}}$ in the COSMOS, XMM-LSS and CDFS fields, respectively. Sources that satisfied a $5\sigma$ peak emission threshold were retained, where $\sigma$ is the r.m.s. at the position of the source, resulting in $28\,267$ detected sources in the COSMOS field, $97\,684$ in XMM-LSS and $17\,866$ in CDFS.

Due to the wide bandwidth of MeerKAT, the effective frequency of the MIGHTEE images varies across each field \citep[see][]{heywood2022}, with a mean value of $\overline{\nu}_{\rm eff}\sim1.28\ {\rm GHz}$. We scale the total flux density of each source, $S_\nu$, from the effective frequency at the source position, obtained from the effective frequency maps provided in the data release, to their values at $1.4\ {\rm GHz}$ by assuming a power-law radio spectrum, $S_\nu\propto\nu^\alpha$, with spectral index $\alpha=-0.7$. We apply a $S_{\rm 1.4\,GHz}>40\ {\rm \mu Jy}$ total flux density cut, at which the source detection completeness is close to one-hundred per cent \citep{hale2025}. The number of radio sources in each field after applying the cut are given in Table~\ref{tab:num_sources}.

\subsubsection{Low- and high-redshift samples}
\label{sssec:high_low_z_samples}

In addition to the angular cross-spectrum between the radio sources and lensing convergence map, we also require measurements of spectra that are sensitive to different redshifts in order to break the degeneracy between the AGN and SFG bias (see Section~\ref{sssec:bias_models}). While samples with different redshift distributions could be constructed by applying redshift cuts to sources in the cross-matched catalogues, the relatively small areas of the regions overlapping with multi-wavelength data result in small sample sizes and large statistical uncertainties in the measured spectra.

\begin{figure}
	\includegraphics[width=\columnwidth]{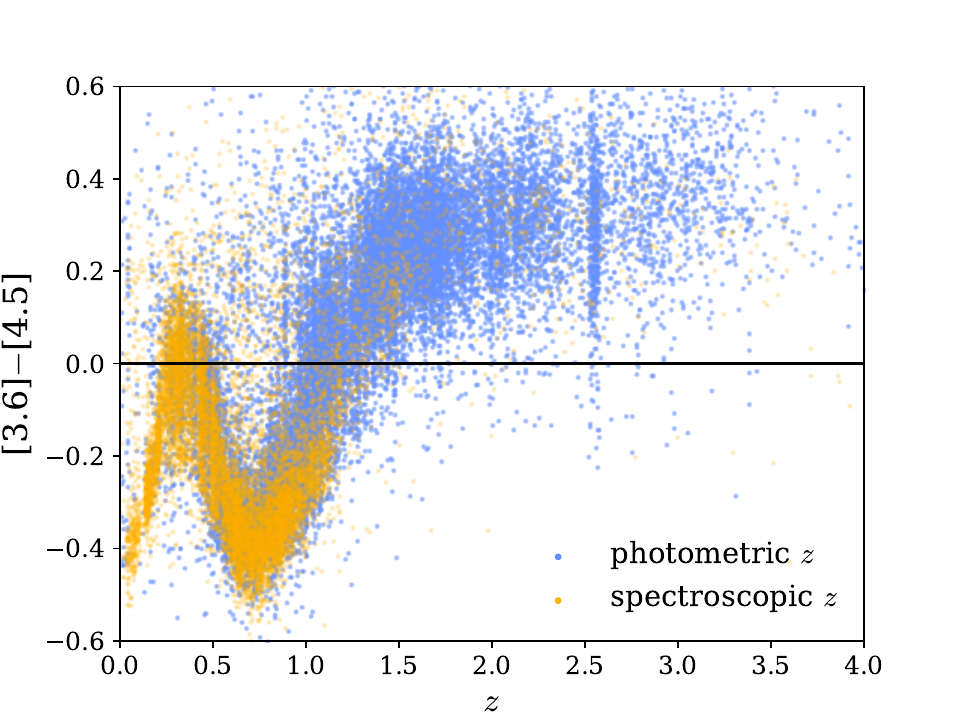}
    \caption{$3.6-4.5\ {\rm \mu m}$ colour against redshift for radio sources cross-matched with \textit{Spitzer} sources that overlap with the multi-wavelength data. Orange points are sources with spectroscopic redshifts, and blue points are the peak redshift PDF values for sources with photometric data. The black horizontal line is the divide used to select the low- and high-redshift samples.}
    \label{fig:colour_redshift}
\end{figure}

Instead, we construct a low- and high-redshift sample by cross-matching with infrared data from the \textit{Spitzer Space Telescope}. In the absence of full photometric redshift information, colours between the IRAC channel 1 and 2 bands ($3.6$ and $4.5\ {\rm \mu m}$) can be used as a crude redshift indicator \citep{papovich2008}. This is because nearly all stellar populations show a peak in their emission at $1.6\ {\rm \mu m}$ due to the minimum in the $\rm H^-$ opacity in stellar atmospheres \citep{john1988, simpson1999}. As a result, galaxies with $z \lesssim 1$ typically display bluer $[3.6]-[4.5]$ colours, while those with $z \gtrsim 1$ appear redder. A simple colour cut can therefore be used to construct samples that preferentially probe lower and higher redshifts. Figure~\ref{fig:colour_redshift} shows the redshift distribution against colour for cross-matched sources in the area that overlaps with the multi-wavelength data where spectroscopic and photometric redshifts are available.

We cross-match to the \textit{Spitzer} catalogues using the likelihood ratio (LR) method \citep{ruiter1977, sutherland1992}. The LR quantifies the probability that a particular infrared source is the true counterpart of a radio source relative to the probability that they are unrelated. It is defined as
\begin{equation}
\label{eqn:likelihood_ratio}
    {\rm LR} = \frac{q(m) f(r)}{n(m)},
\end{equation}
where $q(m)$ is the expected distribution of true counterparts as a function of magnitude, $f(r)$ is the radial probability distribution as a function of separation between the radio and host positions, and $n(m)$ is the magnitude distribution of the entire infrared catalogue \citep{sutherland1992}. We follow the implementation of \cite{mcalpine2012}, who provide a detailed description of how each term in Equation~\ref{eqn:likelihood_ratio} is calculated, using an existing code\footnote{\href{https://github.com/lmorabit/likelihood_ratio}{https://github.com/lmorabit/likelihood\_ratio}}. Matches are accepted above an LR threshold that is chosen as a balance between cross-matching completeness and the reliability of host identification \citep[see][]{williams2019}.

In the XMM-LSS and CDFS fields, we cross-match to the DeepDrill survey catalogues of \cite{lacy2021}, which reach a $5\sigma$ depth of $m_{\rm 3.6\mu m}=23.1$ (AB) and overlap with $8.4$ and $1.5\ {\rm deg}^2$ of the respective MIGHTEE fields. In the COSMOS field, we use the S-COSMOS catalogue of \cite{sanders2007} to a depth of $m_{\rm 3.6\mu m}=23.9$ which overlaps with $2.3\ {\rm deg}^2$.

Prior to matching, regions of poor-quality infrared data, such as areas affected by bright stars or image artefacts, are masked. For DeepDrill, we use the masks supplied with the catalogue release \citep{lacy2021}. In S-COSMOS, we mask areas within $30''$ of objects that are flagged in the catalogues as being close to the border of the images or close to bright stars. The resulting masks are shown in Appendix~\ref{sec:maps_and_masks}.

To minimize stellar contamination, we remove infrared sources located within $1''$ of any \textit{Gaia} DR3 \citep{gaia2023} object classified as a star with probability greater than $90$ per cent. The matching is then performed using the aperture-corrected $3.6\ {\rm \mu m}$ magnitudes, measured with a $2''$ aperture, with the LR statistic. We successfully cross-match $89$, $85$ and $88$ per cent of radio sources in the areas that overlap with the \textit{Spitzer} data in the COSMOS, XMM-LSS and CDFS fields, respectively.

\begin{figure}
	\includegraphics[width=\columnwidth]{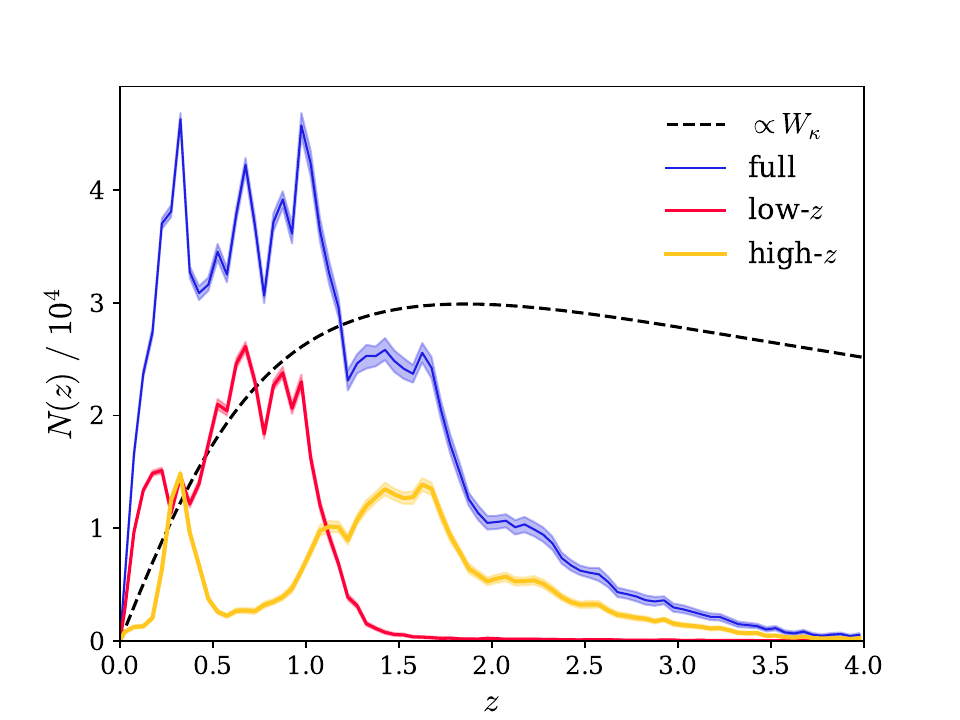}
    \caption{Redshift distributions of radio sources overlapping with the multi-wavelength data (full), and for the low- and high-redshift subsamples (low/high-$z$). The shaded regions show the 16th and 84th percentiles of the redshift distribution realizations. The dashed black line is proportional to the CMB lensing kernel, $W_\kappa$, which peaks at $z\sim1.9$.}
    \label{fig:redshift_distributions}
\end{figure}

After cross-matching, we divide the sample at $[3.6]-[4.5]=0$, which yields roughly the same number of sources on either side of the cut, with the numbers of sources in each sample and the redshifts given in Table~\ref{tab:num_sources}. The corresponding redshift distributions for the low- and high-redshift samples are shown in Figure~\ref{fig:redshift_distributions}. The high-redshift sample contains a small fraction of low-redshift sources, visible in the bump in its redshift distribution at $z \sim 0.4$. We experimented with increasing the $[3.6]-[4.5]$ threshold to $0.1$ and $0.2$ to remove these sources, and found consistent bias parameters but slightly larger uncertainties (the high-redshift sample size is reduced by $\sim60$ per cent when increasing the threshold from $0$ to $0.2$). This low-redshift contribution is taken into account in the modelling through the measured $p(z)$.

\begin{table}
	\centering
	\caption{Effective redshifts (Equation~\ref{eqn:effective_redshift}) and number of sources in each radio sample after applying the $S_{\rm 1.4\,GHz}>40\ {\rm \mu Jy}$ flux density cut. The `Full' sample contains all MIGHTEE sources over the full $\sim20\ {\rm deg}^2$ survey area and is used for the CMB lensing cross-correlation. The `Low-$z$' and `High-$z$' samples are restricted to the smaller MIGHTEE-\textit{Spitzer} overlap regions, and their auto-correlations are computed. The effective redshifts are estimated using the subset of these samples that overlap with the multi-wavelength data.}
	\begin{tabular}{lccccr}
		\hline
		Sample & $z_{\rm eff}$ & COSMOS & XMM-LSS & CDFS & Total\\
		\hline
		Full & 1.11 & $13\,266$ & $45\,914$ & $5\,596$ & $64\,776$ \\
		Low-$z$ & 0.68 & $4\,045$ & $13\,695$ & $2\,722$ & $20\,462$ \\ 
		High-$z$ & 1.46 & $3\,144$ & $13\,346$ & $2\,666$ & $19\,156$ \\
		\hline
	\end{tabular}
    \label{tab:num_sources}
\end{table}

\subsubsection{Redshift distributions}
\label{sssec:redshift_distributions}

Redshift information is obtained from the catalogues of radio sources that have been cross-matched with their near-infrared host galaxies (see \citealt{hale2026} for more information about the cross-matching procedure and catalogues). Host galaxy identification was performed using a combination of the automated \textit{Likelihood Ratio} method \citep[][see Section~\ref{sssec:high_low_z_samples} for further discussion of this method]{ruiter1977, sutherland1992}, and manual host identification by MIGHTEE team members using the MIGHTEE zoo - a \textit{Zooniverse} project \citep{lintott2008}. The regions overlapping with the multi-wavelength coverage span areas of $1.7$, $4.3$ and $1.5\ {\rm deg}^2$ in the COSMOS, XMM-LSS and CDFS fields, respectively. 

Within these overlap regions, spectroscopic redshifts are available for $53$ per cent of cross-matched radio sources in COSMOS, $31$ per cent in XMM-LSS and $22$ per cent in CDFS by matching the hosts with the compiled catalogues of \cite{vaccari2022} within a $1''$ matching radius. For the remaining sources, photometric redshifts are used. In COSMOS and XMM-LSS, photometric redshifts were taken from the catalogues of \cite{hatfield2022} who combined template fitting estimates from \textsc{LePHARE} \citep{arnouts1999, ilbert2006} with the machine learning algorithm GPz \citep{almosallam2016, gomes2018} using a Hierarchical Bayesian model \citep{dahlen2013, duncan2018, duncan2019}. Redshifts for sources not in the catalogues of \citeauthor{hatfield2022} were found using \textsc{LePHARE}. In CDFS, which has less spectroscopic coverage (required to train the machine learning model) and more heterogeneous optical data, photometric redshifts were derived using \textsc{LePHARE} alone (see \citealt{stylianou2026}).

To construct the redshift distribution of each sample while taking the photometric redshift uncertainties into account, we follow a similar approach as \cite{hale2024}. After applying the flux and colour cuts, we sample once from the photometric redshift probability density function (PDF) of each source. For sources with spectroscopic redshifts, we use the measured value. The resulting redshift samples are then binned in a histogram of $\Delta z = 0.05$ up to $z_{\rm max}=4$. We repeat this procedure many times to produce an ensemble of redshift distributions which captures the uncertainty from the photometric redshift PDFs.

Of the radio sources that were successfully cross-matched with a \textit{Spitzer} source, $\sim97$ per cent of those in the multi-wavelength overlap region have an associated redshift. However, $9.2$ per cent of radio sources above the flux density limit in the multi-wavelength overlap regions were not successfully cross-matched. It is plausible that these near-infrared unidentified sources are preferentially located at higher redshift where they are more likely to fall below the survey flux limit. Excluding these sources could therefore bias the inferred redshift distribution and subsequent modelling. This is particularly relevant for the CMB lensing cross-correlation whose kernel remains sensitive to the high redshift tail of the distribution.

To account for these sources, we follow the approach of \cite{thykkathu2026} and construct flux-dependent redshift distributions from the sources that have redshift information. In each field, the sources are divided into radio flux density bins each containing $1000$ sources, and a separate redshift distribution, $p(z|S)$, is constructed within each bin. For each source lacking a host identification, a redshift is then drawn from the $p(z|S)$ corresponding to its radio flux density. We additionally force all sources without a host to be $z>1$ to avoid them being biased to low redshift.

To assess the impact of uncertainty in the redshift distributions on the results, we calculated the theoretical power spectra for each of the redshift distribution realizations. We find that the scatter in the predicted spectra is negligible compared to the measurement uncertainties. We also find that whether or not we impose the $z>1$ restriction of sources without an identified host has a negligible effect on the predicted spectra. We therefore adopt the median normalized redshift distributions across the realizations for the model fitting, under the assumption that the redshift information within the multi-wavelength overlap regions is representative of the whole population. Figure~\ref{fig:redshift_distributions} shows the resulting redshift distributions.

We calculate an effective redshift, $z_{\rm eff}$, for each sample using
\begin{equation}
\label{eqn:effective_redshift}
    z_{\rm eff} = \int {\rm d}z \, z \, p(z).
\end{equation}
The effective redshift values are presented in Table~\ref{tab:num_sources}.

\subsection{ACT DR6 lensing convergence maps}
\label{ssec:act}

We use the ACT DR6 lensing convergence maps\footnote{\href{https://portal.nersc.gov/project/act/dr6_lensing_v1/maps/baseline/}{https://portal.nersc.gov/project/act/dr6\_lensing\_v1/maps/baseline/}} constructed from CMB temperature and polarization data \citep{qu2024, madhavacheril2024, maccrann2024}. The data are available as harmonic coefficients, $\kappa_{\ell m}$, which we transform to a \textsc{HEALPix} \citep{gorski2005} map with resolution parameter $N_{\rm side}$, after truncating $\ell<3N_{\rm side}+1$, using the $\texttt{alm2map}$ function from $\texttt{healpy}$ \citep{zonca2019}. In our analysis, we use $N_{\rm side}=1024$ which corresponds to an angular pixel size of $3.4'$. In Section~\ref{ssec:measurement_robustness_checks} we show that our results are insensitive to the choice of resolution parameter by rerunning our analysis using $N_{\rm side}=512$. The resulting map covers $23$ per cent of the sky and overlaps entirely with the three MIGHTEE fields, which cover just $0.05$ per cent of the sky in total. 


\section{Methods}
\label{sec:methods}

\subsection{Measuring the angular power spectra}

The auto- and cross-spectra are calculated using the pseudo-$C_\ell$ method \citep{peebles1973, hivon2002} as implemented in the \textsc{python} package \textsc{NaMaster}\footnote{\href{https://github.com/LSSTDESC/NaMaster}{https://github.com/LSSTDESC/NaMaster}} \citep{alonso2019}. This requires maps of the fields, $\delta_{\rm g}(\hat{\mathbf{n}})$ and $\kappa(\hat{\mathbf{n}})$, along with their associated sky masks, $w_{\rm g}(\hat{\mathbf{n}})$ and $w_\kappa(\hat{\mathbf{n}})$. Since incomplete sky coverage mixes different spherical harmonic modes, the pseudo-$C_\ell$ approach computes and corrects for the mode-coupling induced by the masks to produce unbiased estimates of the underlying angular power spectra. We bin the measured power spectra in multipole bins with $\Delta \ell = 110$ as a balance between resolution and measurement signal-to-noise.

\subsection{Constructing galaxy overdensity maps}
\label{ssec:galaxy_overdensity_maps}

In this section, the construction of the galaxy overdensity maps, $\delta_{\rm g}(\hat{\mathbf{n}})$, is described. We start by binning the radio sources in a \textsc{HEALPix} map with the same $N_{\rm side}$ as the $\kappa$ map (see Section~\ref{ssec:act}) to create a galaxy counts map, $N_{\rm g}(\hat{\mathbf{n}})$. The observed number density of detected radio sources varies across the footprint of each field due to varying noise, primary-beam response and imaging depth. These effects introduce artificial fluctuations in the source counts that must be accounted for when making the overdensity maps. To remove these observational systematics, we correct the raw number counts of galaxies in each pixel with a position-dependent completeness weight map, $\mathcal{W}_{\rm g}(\hat{\mathbf{n}})$.

The completeness maps are derived from the injection-recovery simulations of \cite{hale2025} that were used to assess the source detection completeness for the MIGHTEE catalogues. They injected simulated radio sources from the \textsc{T-RECS} simulations \citep{bonaldi2019} into the MIGHTEE radio images using realistic source morphology models and taking into account the varying effective frequency of the fields. In each realization, the number of injected sources corresponds to $\sim 10$ per cent of the number of real detected radio sources in the field, and the procedure was repeated for 4000 iterations. We remove injected and recovered sources within $5''$ of real detected radio sources since they would otherwise be re-detected in every iteration and spuriously inflate the completeness. Additionally, sources for which the 50th nearest source in the recovered source catalogue was within the image pixel scale (1.1 arcsec) are removed. These likely reflect source deviations in real, extended sources that result in differences to the numbers of \texttt{PyBDSF} source detections that an object was split into. Not accounting for this resulted in spurious weights due to the over recovery of sources.

The completeness weight in each pixel is the fraction of recovered injected sources:
\begin{equation}
    \mathcal{W}_{\rm g}(\hat{\mathbf{n}}) = \frac{N_{\rm out}(\hat{\mathbf{n}})}{N_{\rm in}(\hat{\mathbf{n}})},
\end{equation}
where $N_{\rm in}(\hat{\mathbf{n}})$ is the number of sources injected into the image at that pixel (after masking around existing sources), and $N_{\rm out}(\hat{\mathbf{n}})$ is the number of those injected sources that were recovered by the source finder with the same flux density cut as applied to the data. 

\begin{figure*}
	\includegraphics[width=2\columnwidth]{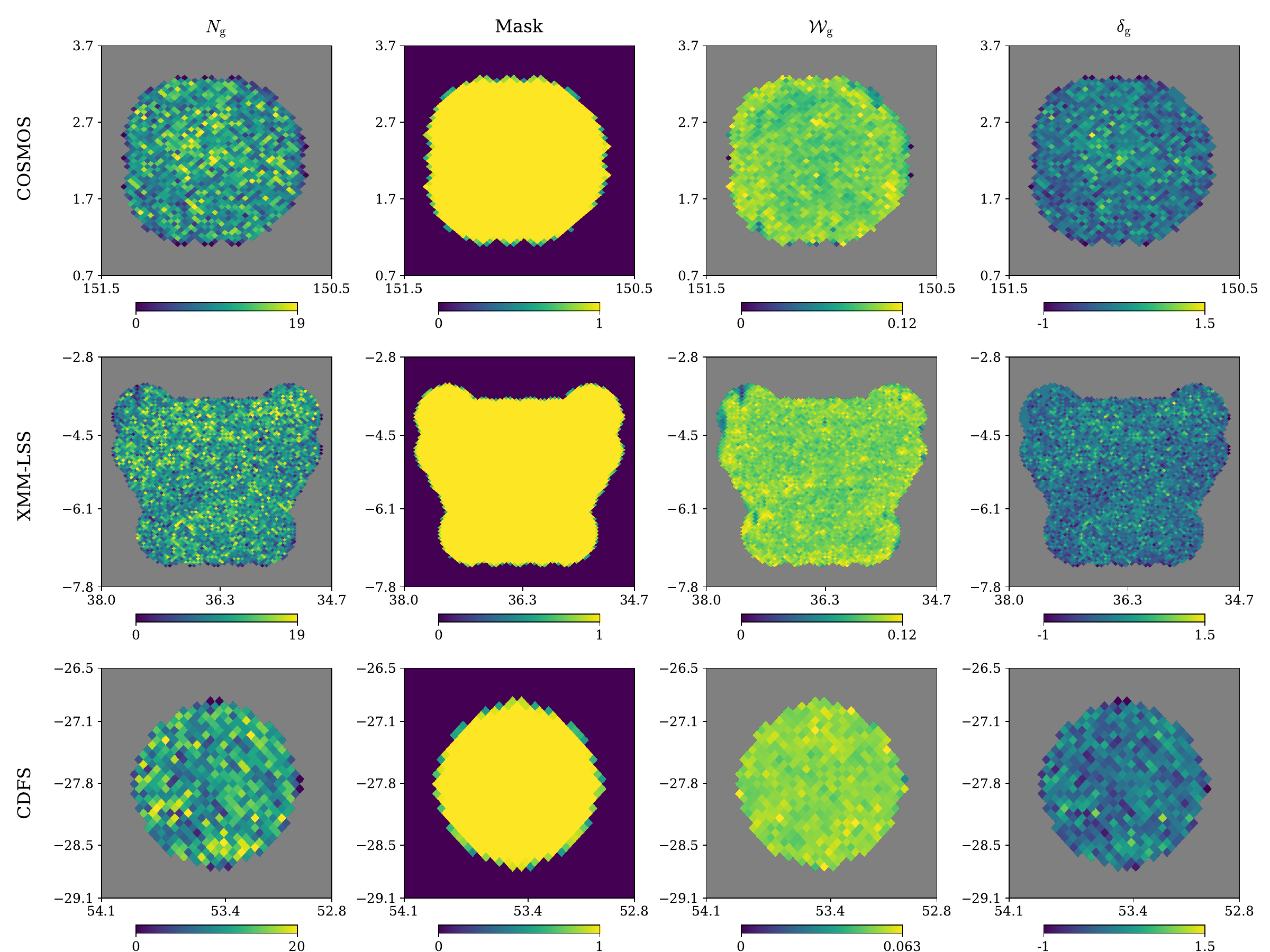}
    \caption{Various maps for the full $S_{\rm 1.4\,GHz} > 40\ {\rm \mu Jy}$ radio sample used in this analysis. From left to right, the columns show the radio galaxy number count maps, $N_{\rm g}$, the masks, the galaxy weights maps, $\mathcal{W}_{\rm g}$, and the galaxy overdensity maps, $\delta_{\rm g}$, for each of the MIGHTEE fields (rows). The numbers on the axes are the right ascensions and declinations of the fields in degrees. All colour bars are linear in scale.}
    \label{fig:maps_and_masks}
\end{figure*}

Taking the weights into account, Equation~\ref{eqn:galaxy_overdensity_definition} becomes
\begin{equation}
    \delta_g(\hat{\mathbf{n}}) = \frac{N_{\rm g}(\hat{\mathbf{n}})}{\overline{N}_{\rm g} \mathcal{W}_{\rm g}(\hat{\mathbf{n}})}-1,
\end{equation}
where the normalization factor $\overline{N}_{\rm g} = \left\langle N_{\rm g}(\hat{\mathbf{n}}) \right\rangle_{\hat{\mathbf{n}}} / \left\langle \mathcal{W}_{\rm g}(\hat{\mathbf{n}}) \right\rangle_{\hat{\mathbf{n}}}$ where $\left\langle \cdots \right\rangle_{\hat{\mathbf{n}}}$ denotes averaging over all pixels within the masks \citep[see e.g.][]{nakoneczny2024}. We set pixels within the lowest $5$ per cent of weights in the mask to zero (along with the corresponding pixels in $\delta_{\rm g}$) to avoid heavily weighted regions. The resulting maps for each MIGHTEE field are shown in Figure~\ref{fig:maps_and_masks}.

\subsection{Field masks and transfer function}
\label{ssec:field_masks}

To obtain unbiased measurements of the angular power spectra, accurate sky masks of the footprint of the data are required. For the cross-correlation between the full radio population and the lensing convergence maps, we make masks for each MIGHTEE field, $w_{\rm g}$, by selecting the area covered by the r.m.s. maps provided in the data release \citep{hale2025}. For the auto-spectra of the low- and high-redshift samples, the catalogue masking described in Section~\ref{sssec:high_low_z_samples} is used. These catalogue footprints are then converted to \textsc{HEALPix} maps with the same resolution as the $\delta_{\rm g}$ and $\kappa$ maps. When downgrading to this resolution, each mask pixel is assigned the fraction of unmasked area contained within that pixel. Any mask pixels with $w_{\rm g}<0.5$ are set to zero to avoid using heavily masked regions. The final masks in each field are shown in Figure~\ref{fig:maps_and_masks}.

For the CMB lensing fields, we use the mask provided in the data release, $w_\kappa$. This was applied to the CMB map prior to lensing reconstruction via the lensing quadratic estimator, so the effective mask entering the reconstructed lensing field involves two powers of $w_\kappa$. The mode-coupling deconvolution in \textsc{NaMASTER}, however, assumes that the mask was applied at the level of the input field. So when computing the cross-correlation, we use $w_{\rm g}$ for mode-coupling with $\delta_{\rm g}$, but $w_\kappa^2$ with $\kappa$ \citep{kim2024}. 

However, this treatment is only approximate because the region selected by $w_{\rm g}$ differs from that used to normalize the lensing reconstruction, leading to small (few per cent) spatially dependent mis-normalizations \citep[for detail, see][]{qu2024}. To account for these residual effects, we follow \cite{kim2024} and estimate a multiplicative transfer function using simulated fields with a known theory spectrum. We use the simulated $\kappa$ maps of \cite{farren2024} who provide $N_{\rm sim}=400$ input convergence maps generated from a fiducial lensing power spectrum. Each input convergence map is used to lens simulated Gaussian CMB temperature and polarization fields which are then processed through the full ACT instrument noise, masking, and lensing reconstruction pipelines to produce corresponding reconstructed convergence maps. 

\begin{figure}
	\includegraphics[width=\columnwidth]{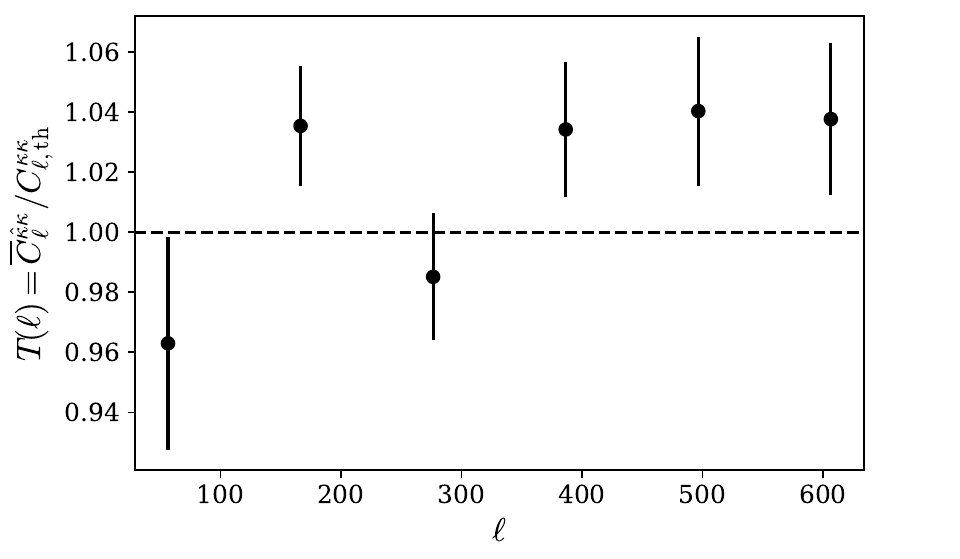}
    \caption{Multiplicative transfer function, $T(\ell)$, used to correct the measured radio-CMB lensing cross-spectrum for residual normalization effects. The error bars show the uncertainty on the mean from the 400 simulations.}
    \label{fig:transfer_function}
\end{figure}

For each simulation, $i$, we compute the cross-correlation of the reconstructed map, $\hat{\kappa}_i$ (masked with $w_\kappa$ and corrected with $w_\kappa^2$ in \textsc{NaMASTER}), with the input convergence field, $\kappa_i$ (masked with $w_{\rm g}$). Averaging over all simulations, we compute the transfer function:
\begin{equation}
    T(\ell) = \frac{\frac{1}{N_{\rm sim}} \sum_{i=1}^{N_{\rm sim}} C_\ell^{\hat{\kappa}_i \kappa_i}}{C_{\ell,\rm th}^{\kappa\kappa}},
\end{equation}
where $C_{\ell,\rm th}^{\kappa\kappa}$ is the predicted theory lensing auto-spectrum (provided with the simulations). The resulting transfer function is shown in Figure~\ref{fig:transfer_function}. To correct the measured cross-spectrum, we simply divide by $T(\ell)$. We do not propagate the uncertainties on $T(\ell)$ as they are much smaller than those on the measured spectra and contribute $\lesssim0.1$ per cent to the total uncertainty.

\subsection{Noise bias}
\label{ssec:noise_bias}

The measured auto-spectra contain additive noise contributions that must be removed to obtain the underlying clustering signal. For the galaxy auto-spectra, the dominant contribution is Poissonian shot-noise arising from the discrete sampling of the density field. This contribution predominantly affects small angular scales (high $\ell$), where shot-noise becomes comparable to or exceeds the clustering signal. In the pseudo-$C_\ell$ formalism, the white (i.e. $\ell$-independent) shot-noise bias is given analytically by 
\begin{equation}
    \label{eqn:noise_bias}
    \tilde{N}_{\ell}^{\rm gg} = \frac{\left\langle w_{\rm g}(\hat{\mathbf{n}}) \right\rangle_{\hat{\mathbf{n}}}}{\overline{N}_\Omega},
\end{equation}
where $\overline{N}_\Omega$ is the average number of galaxies per steradian and $\left\langle w_{\rm g}(\hat{\mathbf{n}}) \right\rangle_{\hat{\mathbf{n}}}$ is the average value of the galaxy mask \citep[i.e., the unmasked fraction of the sky;][]{nicola2020}. This is subtracted from the observed auto-spectra before mode-coupling deconvolution. As a sanity check, we visually verified that the auto-spectra tend to $\tilde{N}_{\ell}^{\rm gg}$ at very high $\ell$.

These noise estimates are conservative since in radio surveys multi-component sources can be identified as separate catalogue entries introducing correlated pairs that enhance the small scale power beyond the Poisson expectation \citep{blake2004}. Moreover, halo exclusion effects \citep{baldauf2016}, as well as stochastic and non-local effects in galaxy formation \citep{kokron2022}, introduce deviations from purely Poissonian noise. Based on \cite{nakoneczny2024} and \cite{tanidis2025}, we therefore marginalize over the noise biases when fitting to the data by fitting it as a free parameter with a $20$ per cent Gaussian prior on its amplitude, $N_\ell^{\rm gg} \to A_{\rm sn} N_\ell^{\rm gg}$.

As discussed later in Section~\ref{sssec:covariance}, computing the covariance matrix requires an estimate of the additive noise bias in the lensing convergence auto-spectrum. The reconstructed $\kappa$ map contains contributions from both the true projected matter field and reconstruction noise arising from instrumental noise, finite CMB modes, and the quadratic estimator procedure \citep{madhavacheril2024}. 

To estimate this bias term, $\tilde{N}_\ell^{\kappa\kappa}$, we use the simulated $\kappa$ maps of \cite{farren2024} (see Section~\ref{ssec:field_masks}). For each simulation, we compute the difference between the reconstructed and input map auto-spectra. Averaging over all simulations yields an estimate for $\tilde{N}_\ell^{\kappa\kappa}$.

\subsection{Model fitting}

\subsubsection{Bias models}
\label{sssec:bias_models}

In addition to measuring the bias of the radio sample, we aim to separate the contributions of AGNs and SFGs, which are expected to trace the underlying matter field differently \citep[e.g.,][]{hale2018, chakraborty2020, mazumder2022}. The effective bias of the radio sample, $b_{\rm radio}(z)$, is a redshift-dependent mixture of the biases of the underlying source populations. We assume that, at a given redshift, the observed radio sources consist of a fraction $f_{\rm AGN}(z)$ of AGNs and a fraction $1-f_{\rm AGN}(z)$ of SFGs. On large scales, where the linear bias applies, the total galaxy overdensity is the number-weighted sum of the overdensities of each population. The effective bias of the combined sample is therefore given by a weighted average of the biases of the two populations:
\begin{equation}
\label{eqn:radio_bias}
    b_{\rm radio}(z) =  f_{\rm AGN}(z)b_{\rm AGN}(z) + \left[ 1-f_{\rm AGN}(z) \right] b_{\rm SFG}(z),
\end{equation}
where $b_{\rm AGN}(z)$, $b_{\rm SFG}(z)$ are the AGN and SFG biases, respectively. Here the AGN bias is treated as a single component, without distinguishing between different radio-AGN sub-populations (HERGs and LERGs, radio-quiet and radio-loud, etc.). 

At a single redshift, or for constant $f_{\rm AGN}$, $b_{\rm SFG}$ and $b_{\rm AGN}$ are completely degenerate; since only the combination $b_{\rm radio}$ is observable, different pairs of $\left\{ b_{\rm AGN}, b_{\rm SFG} \right\}$ that give the same weighted average produce identical predictions for the measured power spectra. However, since $f_{\rm AGN}(z)$ varies with redshift, this degeneracy can be broken by fitting to multiple power spectra that are each sensitive to different redshifts.

To estimate $f_{\rm AGN}(z)$, we use the $1.4\ {\rm GHz}$ radio luminosity functions of \cite{thykkathu2026}, which were derived from the MIGHTEE Early Science data \citep{heywood2022} over $4\ {\rm deg^2}$ of the COSMOS and XMM-LSS fields. We adopt the luminosity functions that were found using their `$z > 1$' method, where radio sources lacking optical identifications (most likely faint galaxies at high redshift) are assigned redshifts $z > 1$ to prevent biasing the analysis towards low-redshift, optically bright populations.

The number density of sources for each population, $p\in\left\{\rm AGN, SFG \right\}$, as a function of redshift is obtained by integrating the luminosity functions, $\Phi_p(L, z)$ (expressed in $\rm Mpc^{-3}\ dex^{-1}$):
\begin{equation}
    n_p(z) = \frac{{\rm d} N_p(z)}{{\rm d} V} = \int_{L_{\rm min}(z)}^\infty \Phi_p(L,z) \, {\rm d} \log_{10}L.
\end{equation}
The lower limit of the integral is the rest-frame luminosity that corresponds to the flux density cut, $S_{\rm cut}$ ($40\ {\rm \mu Jy}$ in this work), at a given redshift:
\begin{equation}
    L_{\rm min}(z) = \frac{4 \pi d_{\rm L}^2(z)}{(1+z)^{1+\alpha}} S_{\rm cut},
\end{equation}
where $d_{\rm L}(z)$ is the luminosity distance. The AGN fraction is then given by:
\begin{equation}
    f_{\rm AGN}(z) = \frac{n_{\rm AGN}(z)}{n_{\rm AGN}(z) + n_{\rm SFG}(z)}.
\end{equation}

For the low- and high-redshift samples, the additional IRAC-based selection could in principle preferentially select AGNs or SFGs, resulting in an AGN fraction that differs from that derived using the luminosity function. We therefore make the additional assumption that the probability of an AGN or SFG entering the low- or high-redshift samples are sufficiently similar such that this selection does not significantly alter the AGN fraction. This assumption is reasonable since the AGN in our sample are predominantly host galaxy dominated rather than quasars, with only of order $\sim200$ genuine quasars over the full survey area \citep{white2025}. The hosts are therefore expected to be dominated by stellar light, as is the SFG population. This means that the selection will be approximately the same - based on the old stellar population.

We investigate four different evolution models for $b_i(z)$ \citep{alonso2021, nakoneczny2024, tanidis2025} where $i \in \left\{ {\rm radio, AGN, SFG} \right\}$:
\begin{itemize}
    \item A simple, non-evolving \textit{constant bias} model where the clustering of galaxies follows the same growth as the matter inhomogeneities (i.e., $\Delta_{\rm g} \propto \Delta_{\rm m}$): 
    \begin{equation}
        b_{i}(z) = b_{0,i}.
    \end{equation}
    \item A \textit{constant amplitude} (or constant clustering) model where the bias is inversely proportional to the linear growth factor:
    \begin{equation}
        b_{i}(z) = \frac{b_{0,i}}{D(z)},
    \end{equation}
    where $D(z) \equiv \Delta_{\rm m}(\mathbf{k},z) / \Delta_{\rm m}(\mathbf{k},0)$ describes the evolution of the matter fluctuations on linear scales ($k\to0$). Here the galaxy overdensity would be a constant amplitude since on linear scales $\Delta_{\rm m}(z) \propto D(z)$.
    \item The \textit{passive} bias evolution model of \citet{fry1996}:
    \begin{equation}
        b_i(z) = 1 + \frac{b_{0,i}-1}{D(z)}.
    \end{equation}
    This model describes a conserved tracer population whose members subsequently evolve only under gravity, with no merging or addition of new tracers. The bias evolves towards unity with time as the tracer overdensity approaches the matter overdensity.
    \item A more flexible, two-parameter \textit{linear} model where the bias increases linearly with redshift:
    \begin{equation}
        b_{i}(z) = b_{0,i} + b_{1,i} z.
    \end{equation}
    This parametrization allows for first-order departures from constant clustering and passive evolution caused by changes in halo occupation or changing source population composition with redshift.
\end{itemize}

\subsubsection{Covariance}
\label{sssec:covariance}

We estimate the analytic Gaussian covariance using \textsc{NaMaster}, following \cite{garcia2019}. This approach models both the CMB lensing convergence and the galaxy overdensities as Gaussian random fields, which has been shown to be a good approximation for the scales studied here \citep{barreira2018, nicola2020}.

The covariance calculation requires an input model for the underlying power spectra. These are obtained using an iterative process following \cite{alonso2021}. First, we compute theoretical spectra, $C_\ell^{ij}$, for all combinations of $i,j \in \left\{ \kappa, {\rm full}, {{\rm low}\text{-}z},{{\rm high}\text{-}z} \right\}$ for a constant bias model with $b_0(z)=1$ using \textsc{CCL}. We then add the appropriate noise contributions (see Section~\ref{ssec:noise_bias}) to each signal. For the galaxy cross-spectra, the shot-noise contribution arises only from sources that are common to both maps. For the cross-correlation between the full and low/high-$z$ radio samples, the surface density term in Equation~\ref{eqn:noise_bias} corresponds to that of the low/high-$z$ sample since these are subsets of the full sample. However, for the cross-correlation of the disjoint low- and high-$z$ samples, no shot-noise term is included. 

\begin{figure}
	\includegraphics[width=\columnwidth]{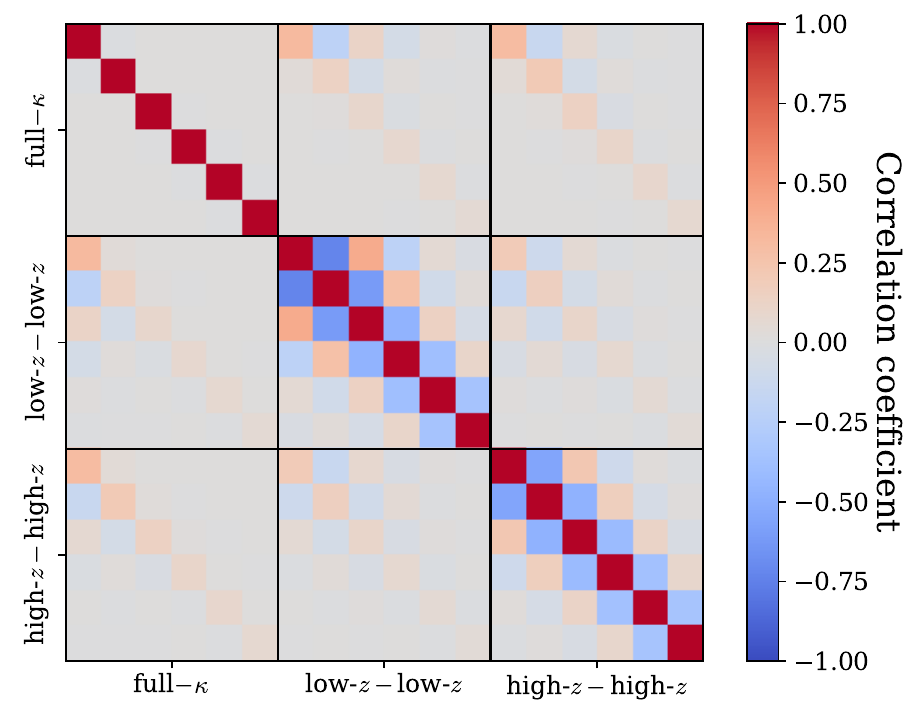}
    \caption{Correlation matrix, ${\rm Corr}_{ij}$, for the cross-spectrum between radio sources and the CMB lensing convergence map (full-$\kappa$), and the auto-spectra of the low- and high-redshift samples.}
    \label{fig:correlation_matrix}
\end{figure}

The resulting covariance matrix is then used in the likelihood (see Section~\ref{sssec:likelihood}) to obtain the best-fitting model parameters. These are then used to compute new theoretical power spectra and produce the final covariance matrix using \textsc{NaMaster}. The correlation matrix, ${\rm Corr}_{ij}={\rm Cov}_{ij}/\sqrt{{\rm Cov}_{ii} {\rm Cov}_{jj}}$, associated with the final covariance matrix is shown in Figure~\ref{fig:correlation_matrix}. We validate the covariance matrix for the CMB cross-spectrum using the ACT simulations in Appendix~\ref{sec:numerical_covariance}.

\subsubsection{Likelihood and fitting method}
\label{sssec:likelihood}

We determine the bias parameters with Bayesian inference using the nested sampling method \citep{skilling2004} as implemented in the \textsc{nautilus}\footnote{\href{https://github.com/johannesulf/nautilus/}{https://github.com/johannesulf/nautilus/}} \citep{nautilus} code, using 1000 live sampling points for each fit. We compare the models by computing the Bayesian evidence, $\mathcal{Z}$, which quantifies the probability of the data, $\mathbf{d}$, given the model, $M$, after marginalizing over the parameter space:
\begin{equation}
\label{eqn:bayesian_evidence}
    \mathcal{Z} = \int {\rm d}\boldsymbol{\theta} \, \mathcal{L}(\mathbf{d}|\boldsymbol{\theta},M) \pi(\boldsymbol{\theta}|M),
\end{equation}
where $\mathcal{L}(\mathbf{d}|\boldsymbol{\theta},M)$ is the likelihood of the data given the model parameters, $\boldsymbol{\theta}$, and $\pi(\boldsymbol{\theta}|M)$ is a prior on those parameters.
The advantage of using a nested sampling method, over MCMC for example, is that the whole parameter space is explored, making the evaluation of Equation~\ref{eqn:bayesian_evidence} straightforward. The uncertainty in $\ln(\mathcal{Z})$ is approximated by $\sigma_{\ln(\mathcal{Z})} \approx 1/\sqrt{N_{\rm eff}}$ where $N_{\rm eff}$ is the effective sample size  from \textsc{nautilus} \citep{nautilus}. We compare the support for different bias evolution models by computing the Bayes factor \citep[e.g.][]{trotta2008}, $\mathcal{B}_{\rm 12} = \mathcal{Z}_{1} / \mathcal{Z}_{2}$, and interpret it using the Jeffreys scale \citep{jeffreys1961}, where $\ln(\mathcal{B}_{12}) > 1,\ 2.5,\ 5$ indicate weak, moderate and strong preference for a model, respectively.

The likelihood is taken to be Gaussian and given by
\begin{equation}
    \ln(\mathcal{L}(\mathbf{d}|\boldsymbol{\theta},M)) = -\frac{1}{2} \chi^2 = -\frac{1}{2} (\mathbf{d}-\mathbf{t}(\boldsymbol{\theta},M))^{\rm T} {\rm Cov^{-1}} (\mathbf{d}-\mathbf{t}(\boldsymbol{\theta},M)),
\end{equation}
where the data vector consists of the observed spectra, $\mathbf{d} = \left( \hat{C}_\ell^{{\rm full}, \kappa}, \hat{C}_\ell^{{\rm low}\text{-}z, {\rm low}\text{-}z}, \hat{C}_\ell^{{\rm high}\text{-}z, {\rm high}\text{-}z} \right)$, and $\mathbf{t}(\boldsymbol{\theta},M)$ is the \textsc{CCL} theoretical prediction for $\mathbf{d}$. 

For the constant bias and constant amplitude models, we adopt flat priors for the model parameters, with $b_{0,i}\in[0,20]$. For the passive evolution model, we require the bias remains physical (i.e. positive) over the fitted redshift range, and therefore include a lower limit of $b_{0,i}\in[1-D(z_{\rm max}),20]$. For the linear bias evolution model, we allow for both positive and negative bias evolution while requiring the bias to remain positive over the fitted redshift range. We therefore parametrize the model in terms of $b_{0,i}$ and $b_{{\rm max},i}=b_i(z_{\rm max})$, assigning both parameters uniform priors of $[0,20]$. This parametrization provides a straightforward mapping of the priors onto a unit hypercube for nested sampling. The slope is then given by $b_{1,i}=(b_{{\rm max},i}-b_{0,i})/z_{\rm max}$. We adopt the high upper bound to allow the bias at the highest redshift to be fully explored.

Because we fit a linear bias model that is independent of scale, we avoid regimes where non-linear clustering invalidates that approximation and restrict the analysis to modes with comoving wavenumber $k < 0.15\ \mathrm{Mpc}^{-1}$. The corresponding multipole cutoff for each sample is evaluated as $\ell_{\rm max}(z_{\rm eff}) = k_{\rm max}\,\chi(z_{\rm eff})$, where $\chi(z_{\rm eff})$ is the comoving distance to the effective redshift (Equation~\ref{eqn:effective_redshift}) of the sample. Using the effective redshift values in Table~\ref{tab:num_sources}, this gives $\ell_{\rm max}(z_{\rm med}) \approx 570$, $390$ and $680$ for the full, low- and high-redshift samples, respectively. Bandpowers beyond these $\ell_{\rm max}$ values are excluded from the likelihood. Because of the finite bandpower windows, the corresponence between $\ell_{\rm max}$ and $k_{\rm max}$ is not exact, and the retained band powers will contain a contribution from scales with $k>k_{\rm max}$. We test whether these contributions are likely to affect the inferred linear bias in Appendix~\ref{sec:lmax_verification}.

To account for the uncertainty in the radio luminosity function when modelling the AGN and SFG bias separately, we numerically marginalize over the posterior distributions of the luminosity function parameters by drawing 200 sets of parameters from their joint posterior distribution. For each set, we evaluate the theoretical power spectrum and corresponding likelihood, and marginalize by averaging the likelihoods.

Due to the large degeneracies when modelling the bias with separate AGN and SFG components, we additionally perform fits that incorporate an external prior on the SFG bias derived from \cite{hale2025b}. They measured two-point clustering of radio sources with $S_{\rm 144\,MHz} > 200\ {\rm \mu Jy}$ ($\sim41\ {\rm \mu Jy}$ at $1.4\ {\rm GHz}$) from the LoTSS Deep Fields \citep{sabater2021, tasse2021}, classified by \cite{best2023}. They fit to both the auto- and cross-correlations of SFGs and LERGs. We use the SFG bias values that were derived from the cross-correlations using a \textsc{HaloFit} model. We implement this constraint as a Gaussian prior on the SFG bias for each redshift bin, 
\begin{equation}
    \ln(\pi_{\rm SFG}) = -\frac{1}{2} \sum_i \left( \frac{ b_{\rm SFG}(z_i) - b_i}{\sigma_i} \right)^2,
\end{equation}
where $z_i$ is the median redshift of the $i$th bin, and $b_i$ and $\sigma_i$ are the corresponding bias and uncertainty from \cite{hale2025b}. Their highest redshift bin is excluded from the prior as it appears inconsistent with the bias evolution trend defined by the lower redshift measurements, and may be affected by source misclassification at high redshift \citep{hale2025b}.



\section{Results and Discussion}
\label{sec:results}

\subsection{Spectra and detection significance}

\begin{figure}
	\includegraphics[width=\columnwidth]{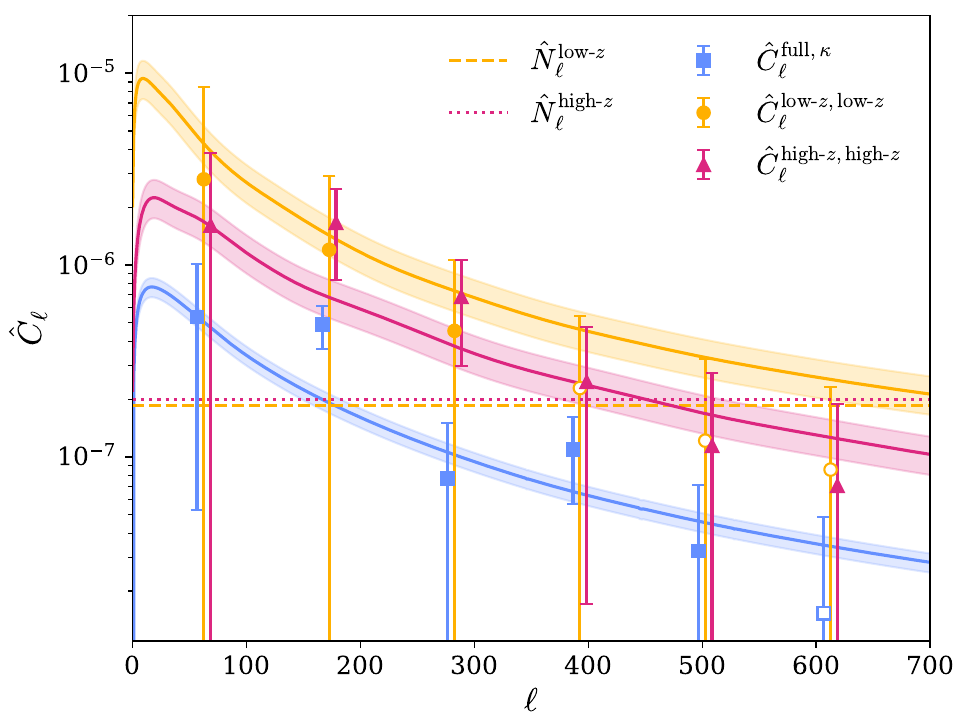}
    \caption{Measured angular power spectrum of the cross-correlation between the full radio sample and the CMB lensing convergence map, $\hat{C}_{\ell}^{{\rm full},\kappa}$ (blue squares), and the auto-spectra of the low- and high-redshift samples, $\hat{C}_{\ell}^{{\rm low}\text{-}z,{\rm low}\text{-}z}$ (orange circles) and $\hat{C}_{\ell}^{{\rm high}\text{-}z,{\rm high}\text{-}z}$ (pink triangles), along with the corresponding noise bias estimates, $\hat{N}_{\ell}^{{\rm low}\text{-}z,{\rm low}\text{-}z}$ (dashed orange line) and $\hat{N}_{\ell}^{{\rm high}\text{-}z,{\rm high}\text{-}z}$ (dotted pink line). Points with hollow markers are above the $\ell_{\rm max}$ value for the sample (see Section~\ref{sssec:likelihood}), and are excluded from the model fitting. The solid lines are the best-fitting theory spectra for the constant amplitude model, and the shaded regions represent the $1\sigma$ uncertainties.}
    \label{fig:measured_power_spectra}
\end{figure}

Figure~\ref{fig:measured_power_spectra} shows the measured cross-spectrum between the full radio sample and CMB lensing convergence, $\hat{C}_{\ell}^{{\rm full},\kappa}$, and the auto-spectra of the low- and high-redshift sub-samples, $\hat{C}_{\ell}^{{\rm low}\text{-}z,{\rm low}\text{-}z}$ and $\hat{C}_{\ell}^{{\rm high}\text{-}z,{\rm high}\text{-}z}$.

In order to establish the statistical significance of the cross-spectrum detection, we fit a single parameter model of the form
\begin{equation}
    \hat{C}_{\ell}^{{\rm full},\kappa} = b C_{\ell}^{{\rm full},\kappa}|_{b_0=1},
\end{equation}
as in \cite{allison2015, alonso2021}. Here, $b$ is a free amplitude parameter and $C_{\ell}^{{\rm full},\kappa}$ is the predicted cross-spectrum for the constant bias model with $b_0=1$. The null hypothesis of no correlation between the galaxy overdensity and CMB lensing convergence maps corresponds to $b=0$. The best-fitting amplitude is determined to be $b=2.61$ with a corresponding $\chi_{\rm best}^2=6.1$ for $\nu=5$ degrees of freedom, whereas the null hypothesis gives $\chi^2_{\rm null}=24.3$. The cross-spectrum is therefore detected with $\sqrt{\chi^2_{\rm null}-\chi^2_{\rm best}} = 4.3\sigma$ statistical significance.

Alternatively, the probability to obtain a $\chi^2$ value greater than $\chi^2_{\rm null}$ given the number of degrees of freedom is $p(\chi^2>\chi^2_{\rm null}) = 2\times10^{-4}$ which corresponds to a $\sim 3.5 \sigma$ detection.

\subsection{Null test}

\begin{figure}
	\includegraphics[width=\columnwidth]{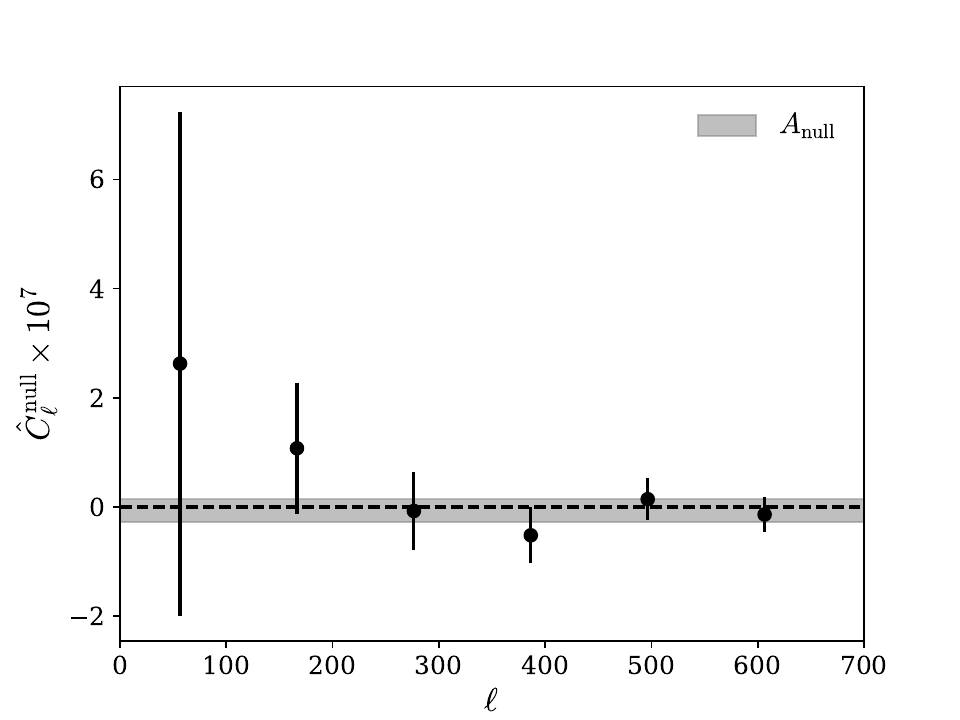}
    \caption{Measured angular power spectrum of the cross-correlation between radio sources and a rotated lensing convergence map. The shaded grey region represents the 1$\sigma$ uncertainties when fitting a constant amplitude, $A_{\rm null}$, to these measurements which is consistent with zero (dashed black line).}
    \label{fig:null_test}
\end{figure}

To validate the cross-spectrum against spurious power from correlated noise, mode-coupling induced by the mask, or other systematics, we perform a null test in which the galaxy overdensity map is cross-correlated with a rotated version of the convergence map. This transformation preserves the noise properties, masking, and overall statistics of the lensing field while removing any true sky correlation between the two maps. The rotation is chosen such that the three MIGHTEE fields remain within the footprint of the rotated map. When the resulting cross-spectrum (Figure~\ref{fig:null_test}) is fitted with a constant amplitude $A_{\rm null}$, we find a best-fitting $A_{\rm null}=\left(-6.1^{+20.1}_{-21.6}\right)\times 10^{-9}$, which is consistent with zero. This means that any detected signal is likely a genuine physical correlation rather than arising from analysis systematics.

\subsection{Measurement robustness tests}
\label{ssec:measurement_robustness_checks}

In this section, we perform a series of sanity checks described in \cite{alonso2021} to assess the stability of the measured power spectra against analysis choices and potential residual systematics.

\begin{figure}
	\includegraphics[width=\columnwidth]{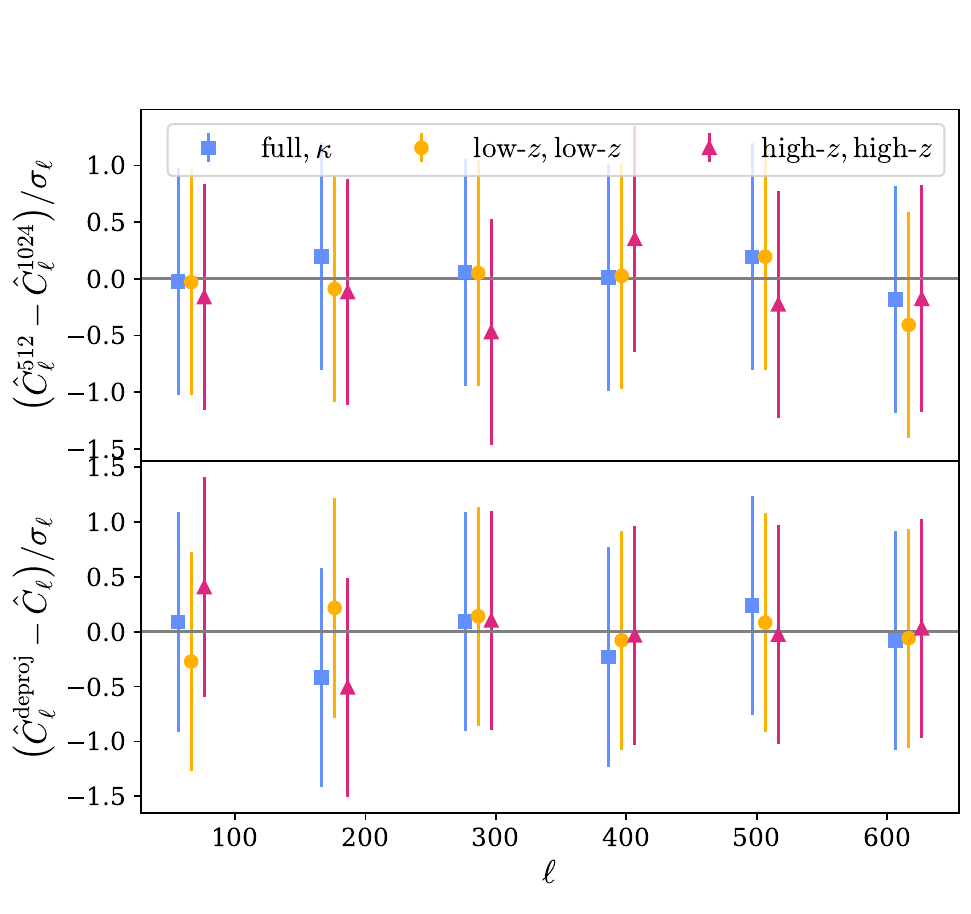}
    \caption{Difference between fiducial and measured angular power spectra for a coarser \textsc{HEALPix} grid with $N_{\rm side}=512$ (top panel) and after deprojecting potential residual noise from the galaxy overdensity maps (bottom panel).}
    \label{fig:sanity_tests}
\end{figure}

First, we verify that the results are robust against the choice of \textsc{HEALPix} map resolution. We remake the maps, masks and power spectra for a coarser grid with half the resolution, $N_{\rm side}=512$. The top panel of Figure~\ref{fig:sanity_tests} shows the relative difference between the original and lower resolution measurements. We find that the results are consistent within the $1\sigma$ uncertainties.

Next, we test for residual contamination from spatial variations in survey depth. Although the galaxy maps are corrected using the completeness weights described in Section~\ref{ssec:galaxy_overdensity_maps}, imperfect modelling of the completeness could introduce excess large-scale power. As a proxy for the effective survey depth, we construct a map proportional to the inverse variance of the r.m.s. noise map in each MIGHTEE field. We then deproject this mean-subtracted template from the galaxy overdensity map when calculating the power spectra following \cite{alonso2019}. The bottom panel of Figure~\ref{fig:sanity_tests} shows that the original and deprojected measurements are again consistent within the uncertainties.

The auto-spectra could be affected by residual systematics or by uncertainties in the shot-noise correction. Before exploring these potential systematics, we first establish a baseline measurement by fitting the low- and high-redshift auto-power spectra individually for the constant bias model. This provides reference galaxy bias values against which the results from the subsequent tests can be compared.

Large-scale systematics, such as residual variations in sensitivity and completeness across individual survey pointings due to the primary beam response, would show up in the power-spectra at low multipoles. For MIGHTEE, the pointing diameter corresponds to $\ell\sim180^\circ/1.4^\circ\approx130$. We therefore repeat the fitting while excluding the lowest multipole bin.

On small scales, the auto-spectra could be affected by a mis-estimation of the noise bias, which was calculated analytically in Section~\ref{ssec:noise_bias}. Because shot-noise is constant across multipoles, any error in its estimation would act as a uniform offset in the measured spectra relative to the true signal. To account for this possibility, we repeat the fits while marginalizing over a constant offset to the observed power spectrum. This is equivalent to modifying the inverse covariance matrix \citep{rybicki1992}:
\begin{equation}
    {\rm Cov}^{-1} \to {\rm Cov}^{-1} - \frac{{\rm Cov}^{-1} \mathbf{1} \mathbf{1}^{\rm T} {\rm Cov}^{-1}}{\mathbf{1}^{\rm T} {\rm Cov}^{-1} \mathbf{1}},
\end{equation}
where $\mathbf{1}$ is a vector of ones with length equal to the number of band powers.

\begin{figure}
	\includegraphics[width=\columnwidth]{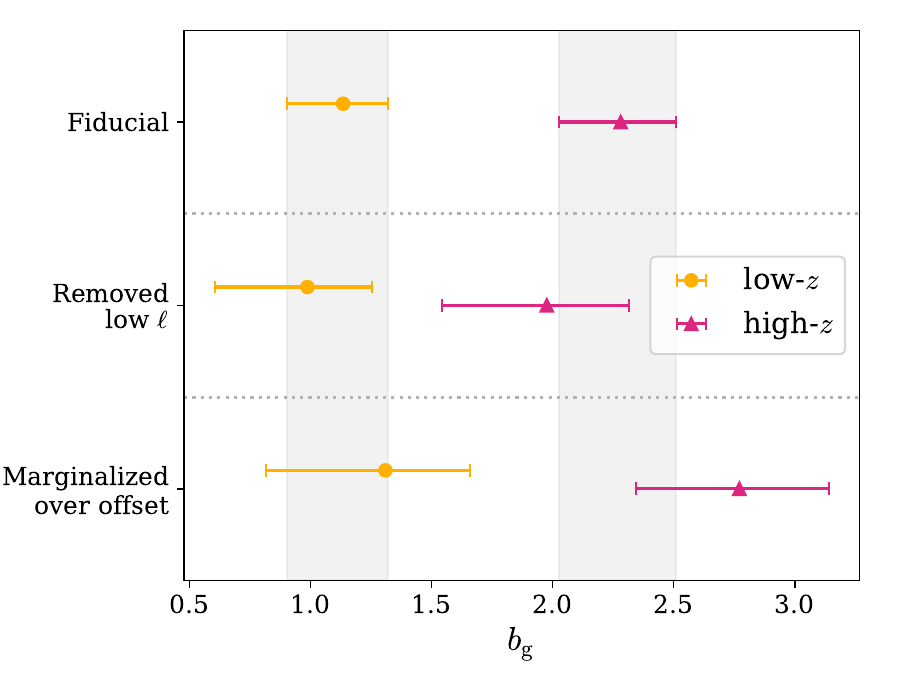}
    \caption{Galaxy bias constraints obtained from fitting to the low- and high-redshift radio auto-power spectra under different systematic tests. The fiducial measurement is compared to fits excluding the lowest multipole bin (to test for large-scale systematics associated with survey pointings) and fits where a constant offset in the power spectrum is marginalized over to account for potential mis-estimation of the shot-noise contribution.}
    \label{fig:auto_spectra_tests}
\end{figure}

The resulting galaxy bias values for the fiducial fit, the low-$\ell$ truncated fit, and the shot-noise-marginalized fit are displayed in Figure~\ref{fig:auto_spectra_tests}. All estimates are consistent within their uncertainties with the fiducial bias value for both auto-spectra, demonstrating that the inferred galaxy bias is stable against both large- and small-scale systematic effects.

\subsection{Effective bias of radio sources}

{\renewcommand{\arraystretch}{1.3}
\begin{table*}
    \centering
    \caption{Best-fitting parameters and auto-spectra shot-noise amplitudes for each bias evolution model without separating the AGN and SFG bias contributions, together with the corresponding reduced $\chi^2$ values and logarithms of the Bayesian evidence. The uncertainties correspond to the 16th and 84th percentiles of the posteriors.}
    \begin{tabular}{lccccccr}
        \hline
        Model & $b_0$ & $b_1$ & $A_{\rm sn}^{\rm low}$ & $A_{\rm sn}^{\rm high}$ & $\chi^2_\nu$ & $\ln(\mathcal{Z})$ \\
        \hline
            Constant bias & ${1.63}^{+0.19}_{-0.22}$ & $-$ & ${0.98}^{+0.20}_{-0.20}$ & ${1.11}^{+0.18}_{-0.16}$ & 1.60 & $-12.81\pm0.01$ \\ 
Constant amplitude & ${1.20}^{+0.12}_{-0.14}$ & $-$ & ${0.98}^{+0.19}_{-0.20}$ & ${1.01}^{+0.18}_{-0.17}$ & 1.13 & $-10.51\pm0.01$ \\ 
Passive & ${1.51}^{+0.13}_{-0.14}$ & $-$ & ${0.96}^{+0.20}_{-0.19}$ & ${1.07}^{+0.18}_{-0.18}$ & 1.39 & $-11.97\pm0.01$ \\ 
Linear & ${0.49}^{+0.44}_{-0.32}$ & ${1.62}^{+0.37}_{-0.51}$ & ${0.99}^{+0.19}_{-0.20}$ & ${0.93}^{+0.18}_{-0.19}$ & 0.95 & $-10.20\pm0.01$ \\ 
        \hline
    \end{tabular}
    \label{tab:radio_bias_parameters}
\end{table*}
}

In this section, we discuss the best-fitting bias model parameters from fitting the effective bias without separating the AGN and SFG contributions. The best-fitting parameters, along with the reduced $\chi^2$ and Bayesian evidence values, are presented in Table~\ref{tab:radio_bias_parameters}.

The best-fitting amplitudes for the galaxy shot-noise are consistent with unity for all models for both the low- and high-$z$ samples. Since we fit to the linear regime, where the contribution from the noise is subdominant to the correlation signal, varying the amplitudes has little impact on the inferred parameters, and we found that fitting the models with fixed amplitudes results in very similar uncertainties on the bias model parameters.

For the constant bias model, we find a best-fitting $b_0=1.63^{+0.19}_{-0.22}$ with an acceptable $\chi^2_\nu=1.60$. The constant amplitude model returns $b_0=1.20^{+0.12}_{-0.14}$ and an improved fit with $\chi^2_\nu=1.13$. Comparing the evidence for these two models gives a Bayes factor $\ln(\mathcal{B}) = \Delta \ln(\mathcal{Z})=2.30$ in favour of the constant amplitude model, so there is weak evidence that this model is preferred, indicating that the data is better described by a bias that increases with redshift.

The passive evolution model of \cite{fry1996} gives $b_0=1.51^{+0.13}_{-0.14}$, intermediate between the values obtained for the constant bias and constant amplitude models. Comparing the evidence gives $\Delta \ln(\mathcal{Z})=1.46$ marginally in favour of the constant amplitude model, providing weak evidence against passive evolution. The \citeauthor{fry1996} model describes a conserved population whose members evolve under gravity with a constant comoving number density. Evolution in the source number density could be driven by mergers or new formation, or from changes in the relative contributions of AGNs and SFGs to the radio population.

The lowest reduced chi-squared value is achieved by the linear model with $\chi^2_\nu=0.95$. However, the evidence difference relative to the constant amplitude model, $\Delta \ln(\mathcal{Z}) = 0.31$, indicates no significant preference for either model. This suggests that the improvement in goodness-of-fit is not sufficient to justify the additional model parameter.

\begin{figure}
	\includegraphics[width=\columnwidth]{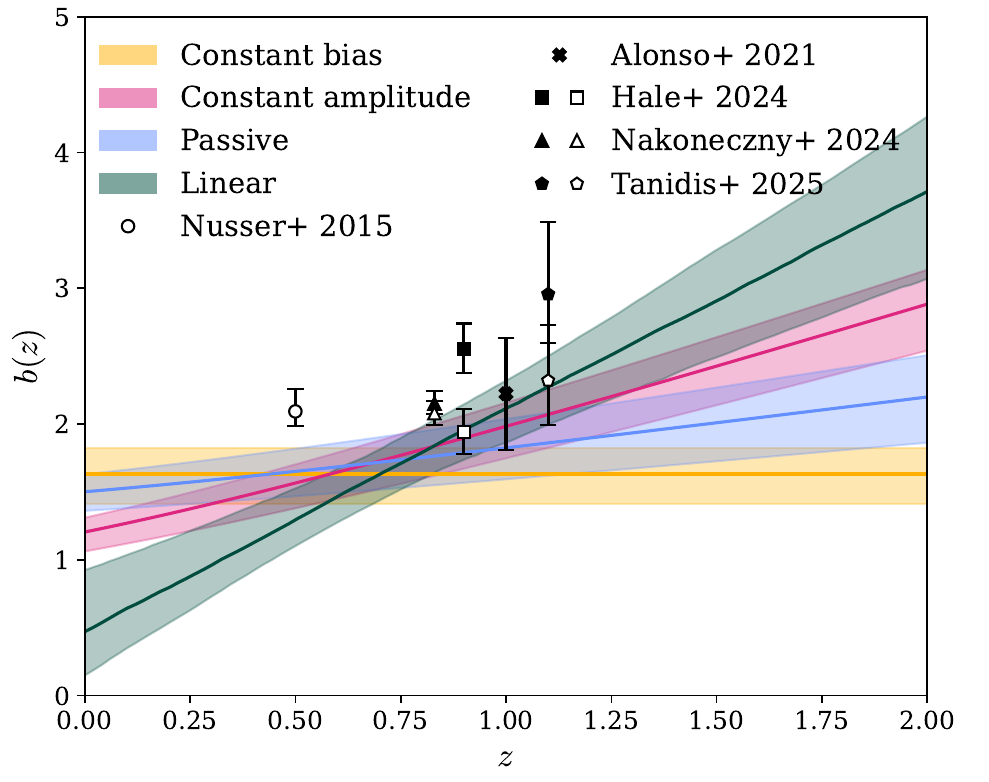}
    \caption{Best-fitting bias evolution for the combined radio sample (i.e. without separating AGN and SFG contributions) for the four models considered: constant bias (yellow), constant amplitude (pink), passive (blue) and linear (green). Solid curves show the posterior median predictions, with shaded bands indicating the $1\sigma$ uncertainties. Also shown are previous measurements for mixed radio populations from \protect \cite{nusser2015}, \protect \cite{alonso2021}, \protect \cite{hale2024}, \protect \cite{nakoneczny2024}, and \protect \cite{tanidis2025}. Filled (black) markers denote results obtained with a constant amplitude model, while open (white) markers correspond to constant bias fits.}
    \label{fig:radio_bias}
\end{figure}

Figure~\ref{fig:radio_bias} shows the evolution of the best-fitting bias models against redshift. The linear model favours a steeper slope than the constant amplitude model with a gradient $b_1 = 1.62^{+0.37}_{-0.51}$ and a low normalization $b_0 = 0.49^{+0.44}_{-0.32}$. The constant amplitude model corresponds to a galaxy overdensity that remains fixed on linear scales, so interpreted physically, a steeper evolution corresponds to a radio population whose overdensity decreases towards later times. This suggests that the radio sources at high redshift inhabit rarer and more massive dark matter haloes. However, this evolution may also be explained by the changing population mix with redshift; the AGN fraction increases towards higher redshift, and the flux limit corresponds to a progressively higher luminosity threshold which will preferentially select AGN. Since AGNs are known to inhabit more clustered environments than SFGs \citep[e.g.,][]{hale2018, chakraborty2020, mazumder2022}, the steep bias evolution inferred by the linear model likely reflects this changing population composition rather than intrinsic evolution of the environments of a single population of radio galaxies.

Also shown in Figure~\ref{fig:radio_bias} are previous measurements of the bias for mixed radio populations, spanning a range of frequencies and flux limits. These include: the bias values of \cite{nusser2015} derived from the auto-spectrum of sources in NVSS \citep{condon1998}; cross- and auto-spectrum fits using the first \citep{alonso2019} and second \citep{nakoneczny2024} data releases of LoTSS \citep{vanhaarlem2013, shimwell2019, shimwell2022}; the auto-correlation of LoTSS second data release sources from \cite{hale2024}; and the auto- and cross-spectrum of sources from EMU \citep{norris2011, norris2021} by \cite{tanidis2025}.

Our measurements are broadly consistent with previous studies, with all results agreeing within the $2\sigma$ uncertainties. The bias values inferred under the constant bias model are slightly lower than those reported by \citeauthor{nusser2015}, \citeauthor{nakoneczny2024}, and \citeauthor{tanidis2025}, while the constant amplitude model yields values lower than \citeauthor{hale2024} and \citeauthor{tanidis2025}. This difference is likely due to our lower flux density threshold compared to those used in these studies, which will preferentially select a population with a smaller AGN fraction and consequently a lower effective bias. For reference, Table~\ref{tab:previous_flux_limits} lists the flux limits of these works converted to an equivalent value at $1.4\ {\rm GHz}$ by assuming $S_{\nu} \propto \nu^{-0.7}$.

{\renewcommand{\arraystretch}{1}
\begin{table}
    \centering
    \caption{Flux density limits of previous bias measurements for mixed radio populations. The second and third column list the flux limit, $S_{\nu, {\rm lim}}$, and frequency, $\nu$, at which it was applied in each study. The fourth column gives the corresponding flux limit scaled to $1.4\ {\rm GHz}$ by assuming $S_{\nu} \propto \nu^{-0.7}$.}
    \begin{tabular}{lcccr}
        \hline
        Reference & $S_{\nu, {\rm lim}}$ & $\nu$ & $S_{{\rm 1.4\,GHz}, \, {\rm lim}}$ \\
        & [$\mu$Jy] & [MHz] & [$\mu$Jy] \\
        \hline
            \cite{nusser2015} & 7\,200 & 1\,400 & 7\,200 \\
            \cite{alonso2021} & 2\,000 & 144 & 410 \\
            \cite{hale2024} & 1\,500 & 144 & 310 \\
            \cite{nakoneczny2024} & 1\,500 & 144 & 310 \\
            \cite{tanidis2025} & 180 & 944 & 140 \\
            This work & 40 & 1\,400 & 40 \\
        \hline
    \end{tabular}
    \label{tab:previous_flux_limits}
\end{table}
}

\subsection{Bias of AGNs and SFGs}
\label{ssec:agn_sfg_bias}

{\renewcommand{\arraystretch}{1.3}
\begin{table*}
    \centering
    \caption{Best-fitting bias model parameters and auto-spectra shot-noise amplitudes when separating the AGN and SFG contributions to the total effective bias, together with the corresponding reduced $\chi^2$ values and logarithms of the Bayesian evidence. The top rows show the results when fitting without using the SFG bias values of \protect \cite{hale2025b} as priors, and the bottom rows show the results when including this prior. The uncertainties correspond to the 16th and 84th percentiles of the posteriors.}
    \begin{tabular}{lcccccccc}
        \hline
        Model & $b_{0, {\rm AGN}}$ & $b_{0, {\rm SFG}}$ & $b_{1, {\rm AGN}}$ & $b_{1, {\rm SFG}}$ & $A^{\rm low}_{\rm sn}$ & $A^{\rm high}_{\rm sn}$ & $\chi^2_\nu$ & $\ln(\mathcal{Z})$ \\
        \hline Without SFG prior \\ \hline
Constant bias & ${5.34}^{+0.94}_{-1.35}$ & ${0.41}^{+0.52}_{-0.30}$ & $-$ & $-$ & ${0.97}^{+0.20}_{-0.19}$ & ${1.00}^{+0.18}_{-0.19}$ & 1.17 & $-11.46\pm0.01$ \\ 
Constant amplitude & ${2.60}^{+0.65}_{-0.91}$ & ${0.52}^{+0.47}_{-0.34}$ & $-$ & $-$ & ${0.98}^{+0.20}_{-0.20}$ & ${0.92}^{+0.19}_{-0.18}$ & 0.95 & $-11.02\pm0.01$ \\ 
Passive & ${2.83}^{+0.50}_{-0.70}$ & ${0.98}^{+0.29}_{-0.17}$ & $-$ & $-$ & ${0.97}^{+0.20}_{-0.20}$ & ${0.97}^{+0.18}_{-0.18}$ & 1.13 & $-12.28\pm0.01$ \\ 
Linear & ${1.65}^{+1.84}_{-1.18}$ & ${0.42}^{+0.47}_{-0.30}$ & ${1.50}^{+1.23}_{-1.27}$ & ${0.76}^{+1.03}_{-0.65}$ & ${0.99}^{+0.20}_{-0.20}$ & ${0.92}^{+0.18}_{-0.19}$ & 1.33 & $-11.73\pm0.01$ \\ 
\hline With SFG prior \\ \hline
Constant bias & ${3.71}^{+0.72}_{-0.79}$ & ${1.11}^{+0.08}_{-0.08}$ & $-$ & $-$ & ${0.97}^{+0.20}_{-0.19}$ & ${1.04}^{+0.17}_{-0.18}$ & 1.43 & $-9.81\pm0.02$ \\ 
Constant amplitude & ${1.96}^{+0.37}_{-0.40}$ & ${0.90}^{+0.06}_{-0.06}$ & $-$ & $-$ & ${0.98}^{+0.20}_{-0.20}$ & ${0.96}^{+0.18}_{-0.19}$ & 1.06 & $-8.57\pm0.02$ \\ 
Passive & ${2.63}^{+0.39}_{-0.42}$ & ${1.11}^{+0.06}_{-0.06}$ & $-$ & $-$ & ${0.96}^{+0.20}_{-0.19}$ & ${0.97}^{+0.19}_{-0.18}$ & 1.19 & $-9.16\pm0.02$ \\ 
Linear & ${1.30}^{+1.36}_{-0.92}$ & ${0.58}^{+0.18}_{-0.18}$ & ${1.38}^{+0.92}_{-1.16}$ & ${1.25}^{+0.38}_{-0.39}$ & ${0.98}^{+0.19}_{-0.20}$ & ${0.93}^{+0.18}_{-0.18}$ & 1.23 & $-8.39\pm0.02$ \\
        \hline
    \end{tabular}
    \label{tab:agn_sfg_bias_parameters}
\end{table*}
}

\begin{figure*}
	\includegraphics[width=2\columnwidth]{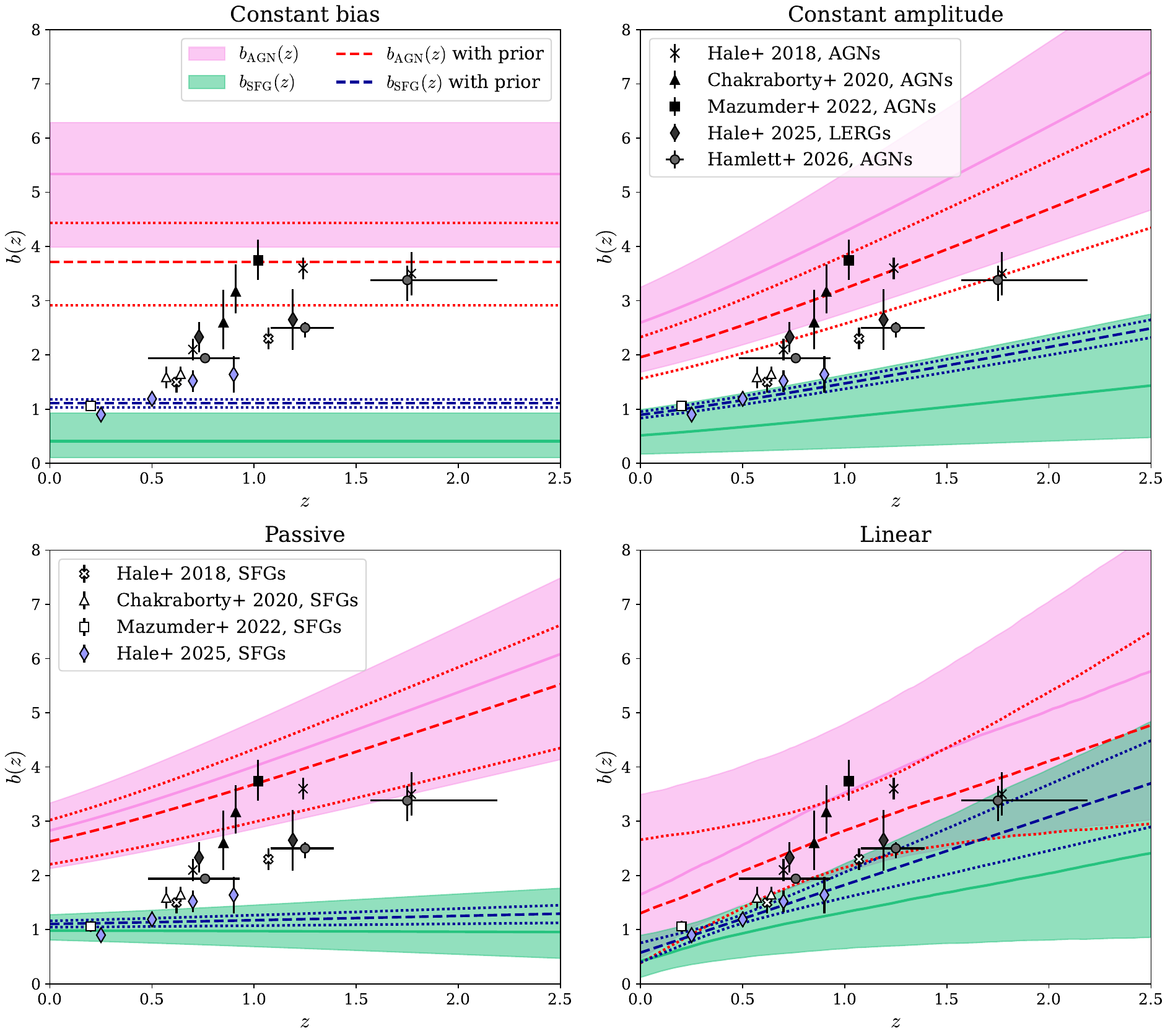}
    \caption{Best-fitting bias evolution for AGNs (pink and red) and SFGs (green and blue) for the four models considered: constant bias (top left), constant amplitude (top right), passive (bottom left) and linear (bottom right). Solid curves show the posterior median predictions, with shaded bands indicating the $1\sigma$ uncertainties, when fitting without using the SFG bias values of \protect \cite{hale2025b} (blue diamonds) as a prior. The dashed and dotted lines are the median posterior predictions and $1\sigma$ uncertainties from fitting with the prior. Also shown are the bias values for LERGs (dark grey diamonds) from \protect \cite{hale2025b} and AGNs (grey circles) from \protect \cite{hamlett2026}, along with the AGN and SFG bias measurements from \protect \cite{hale2018} (crosses), \protect \cite{chakraborty2020} (triangles) and \protect \cite{mazumder2022} (squares). Dark grey and black markers denote AGN bias measurements, while white markers denote SFG bias measurements.}
    \label{fig:agn_sfg_bias}
\end{figure*}

We now discuss the best-fitting bias model parameters when the AGN and SFG contributions to the total effective bias are modelled separately. Table~\ref{tab:agn_sfg_bias_parameters} presents the best-fitting bias model parameters, together with reduced $\chi^2$ values and Bayesian evidence. We fit to the observed power spectra both with and without imposing the SFG bias measurements of \cite{hale2025b} as a prior on $b_{\rm SFG}(z)$.

Figure~\ref{fig:agn_sfg_bias} shows the four bias evolution models alongside the SFG and LERG bias measurements from \cite{hale2025b}, as well as the AGN bias values reported by \cite{hamlett2026} who measured the cross-correlation between AGNs from MIGHTEE and a large population of near-infrared-selected galaxies. Also shown are the AGN and SFG bias measurements of \cite{hale2018}, who measured the clustering of classified radio sources in the COSMOS field using VLA observations at $3\ {\rm GHz}$ \citep{smolcic2017}; \cite{chakraborty2020}, who used $400\ {\rm MHz}$ and $612\ {\rm MHz}$ archival data from the upgraded Giant Metrewave Radio Telescope \citep[uGMRT;][]{chakraborty2019}; and \cite{mazumder2022} who investigated clustering at $325\ {\rm MHz}$ using GMRT data \citep{swarup1991}. 

Without the prior, the inferred AGN bias for the constant bias model, $b_{0, \rm AGN} = 5.34^{+0.94}_{-1.35}$, lies significantly above the values reported by \cite{hamlett2026}, while the SFG bias, $b_{0, \rm SFG} = 0.41^{+0.52}_{-0.30}$ is only consistent with the lowest redshift point of \cite{hale2025b}. Modelling the AGN and SFG components separately yields $\chi^2_\nu=1.17$, an improvement over the single radio bias model ($\chi^2_\nu = 1.60$). This is because the effective bias of the model changes as the AGN fraction evolves with redshift (see Equation~\ref{eqn:radio_bias}), better capturing the evolution of the clustering amplitude.

The constant amplitude model provides the best overall fit ($\chi^2_\nu=0.95$) with $b_{0, \rm AGN} = 2.60^{+0.65}_{-0.91}$ and $b_{0, \rm SFG} = 0.52^{+0.47}_{-0.34}$. However, now that the constant bias model also produces an evolving effective bias, the evidence no longer shows a preference for the constant amplitude model, with $\Delta \ln(\mathcal{Z}) = 0.44$. AGNs are found to be more biased tracers than SFGs at all redshift, in agreement with the findings of previous work \citep{magliocchetti2017, hale2018, chakraborty2020, mazumder2022}. The resulting SFG bias is consistent with the measurements of \cite{mazumder2022}, \cite{hale2025b} and the low-redshift measurement of \cite{hale2018}, although it is lower than their higher-redshift measurement, and those of \cite{chakraborty2020}. The AGN bias is in agreement with previous results, though it remains high relative to \cite{hamlett2026}. Again, modelling the AGN and SFG contributions separately represents an improvement in the goodness-of-fit compared to a single radio bias model ($\chi^2_\nu=0.95$ compared to $1.13$), showing that the evolving population mix is an important component of the observed clustering evolution.

The passive evolution model gives $b_{0, {\rm AGN}} = 2.83^{+0.50}_{-0.70}$ and $b_{0, {\rm SFG}} = 0.98^{+0.29}_{-0.17}$. In the \cite{fry1996} model, the bias evolves towards unity at late times, such that $b_0>1$ corresponds to a bias that increases with redshift, while $b_0<1$ corresponds to a decrease. The posterior for $b_{0, {\rm SFG}}$ therefore permits both forms of evolution at the $1\sigma$ level, with the median value corresponding to a mild decrease with redshift. This is in contrast to the trend seen in the previous SFG bias measurements (Figure~\ref{fig:agn_sfg_bias}), which show a steeper increase towards high redshift. The constant amplitude model is weakly favoured by the evidence, with $\Delta \ln(\mathcal{Z}) = 1.26$, suggesting that AGNs and SFGs are likely not conserved tracers.

The linear model yields large parameter uncertainties and the worst fit with $\chi^2_\nu = 1.33$. As can be seen in Figure~\ref{fig:agn_sfg_bias}, the resulting AGN and SFG bias are, however, consistent with all previous measurements within the large uncertainties.

When the SFG prior from \cite{hale2025b} is imposed, the constant bias model shifts the AGN bias normalization into agreement with the highest redshift bin of \cite{hamlett2026}. The corresponding SFG normalization becomes higher but still within the $\sim1\sigma$ uncertainties of the fit without the prior ($b_{0, \rm SFG} = 1.11^{+0.08}_{-0.08}$ compared to $0.41^{+0.52}_{-0.30}$). This model is, however, mildly disfavoured by the evidence relative to the constant amplitude model ($\Delta \ln(\mathcal{Z})=1.24$).

As with the effective radio bias, the gradients of the best-fitting linear models, both with and without the SFG prior, are steeper than the evolution of the constant amplitude models, implying that the inferred overdensities decrease towards later times. This is contrary to studies that find evidence for an increase in the AGN dark matter halo mass over cosmic time \citep{hale2018, hamlett2026}. One plausible explanation for this is luminosity driven selection; a fixed flux limit will preferentially select more luminous sources at higher redshift, and there is substantial evidence that more luminous radio-AGNs inhabit more strongly clustered environments \citep{bardelli2010, lindsay2014, hale2018, croston2019, mo2020}. The steep AGN bias evolution may therefore reflect the changing luminosity rather than genuine environmental evolution of a fixed population. 

Rapid evolution in the environments of SFGs has also been observed in previous work. \cite{hale2025b} found a steeper evolution in the SFG bias relative to the linear growth factor (see the top right panel of Figure~\ref{fig:agn_sfg_bias}). Compiling results from previous studies, \cite{magliocchetti2014} found that, regardless of selection criteria, the dark matter haloes of SFGs at $z\gtrsim1.5$ are $\sim1.5\ {\rm dex}$ more massive than those in the local Universe. This may also explain why our SFG bias assuming the constant amplitude evolution is lower than the high-redshift measurement from \cite{hale2018}. While our steep gradient may support these findings, the linear model cannot be distinguished from the constant amplitude model based on the Bayesian evidence ($\Delta \ln(\mathcal{Z}) = 0.71$). A larger sample with smaller statistical uncertainties on the clustering signal is therefore required to determine whether the bias evolution departs significantly from constant clustering evolution.

Since $b_{1, \rm AGN}$ and $b_{1, \rm SFG}$ are consistent, we also fit the linear model but with the gradients of the AGN and SFG components tied by fixing $b_{1, \rm AGN}=b_{1, \rm SFG}$, reducing the number of free parameters. The resulting bias parameters and best-fitting model are presented in Appendix~\ref{sec:linear_tied}. When including the SFG prior, we find a best-fitting gradient of $b_1=1.31^{+0.30}_{-0.33}$, which lies between the values obtained when the AGN and SFG gradients are allowed to vary independently. This model has the largest Bayesian evidence ($\ln(\mathcal{Z}) = -7.89$), reflecting the reduced model complexity, but is still not favoured relative to the constant amplitude and linear models. Moreover, tying the gradients of the AGN and SFG bias evolution is not motivated physically, so we do not consider this model further.

\cite{hamlett2026} selected AGNs using a luminosity threshold corresponding to an approximate flux limit of $S_{\rm 1.4\,GHz} \gtrsim 130\ {\rm \mu Jy}$. If radio bias increases with luminosity, we may therefore expect our sample to yield a lower AGN bias. However, the AGN population mix must also be considered. The fraction of HERGs is known to increase with both radio luminosity and redshift \citep{best2012, pracy2016, whitam2016, whittam2018, best2023}, while LERGs typically reside in denser and more strongly clustered environments \citep{hardcastle2004, ineson2015, ching2017, hale2018, lopez2026}. When compared to the LERG bias measurements of \cite{hale2025b}, the AGN bias inferred from our constant amplitude model is systematically higher, but is consistent within the uncertainties. This suggests that the relatively high AGN bias we recover may reflect a larger LERG contribution within our sample.

To separate the contributions from AGNs and SFGs to the total bias, we used the radio luminosity functions of \cite{thykkathu2026} to calculate the AGN fraction as a function of redshift. These luminosity functions have a steeper faint-end slope than some previous works \citep{novak2017, novak2018}, and there is evidence that they over predict the low-redshift source counts at the faintest fluxes \citep{hale2026}. We therefore repeated the analysis using luminosity functions derived from $3\ {\rm GHz}$ VLA-COSMOS data \citep{novak2018}, which have a flatter faint-end slope. The latter predicts fewer SFGs at low redshift, leading to a higher inferred AGN fraction. Consequently, a lower AGN bias is required to reproduce the same observed total bias. The resulting model parameters, obtained using this alternative luminosity function, are presented in Table~\ref{tab:agn_sfg_bias_parameters_alternative_luminosity_function}. Although the inferred bias parameters are systematically lower than those in Table~\ref{tab:agn_sfg_bias_parameters}, the shifts are well within the statistical uncertainties and would not affect any of the conclusions or interpretation of the results.

\section{Conclusions}
\label{sec:conclusions}

In this work, we measured the harmonic-space cross-correlation power spectrum between the projected overdensity field of $S_{\rm 1.4\,GHz}>40\ {\rm \mu Jy}$ radio sources from MIGHTEE DR1 and the CMB lensing convergence field from ACT DR6. We detect the cross-spectrum at a significance of $\sim4\sigma$. By cross-matching the radio catalogue with \textit{Spitzer} galaxies from the S-COSMOS and DeepDrill surveys, we constructed low- and high-redshift samples using a $3.6-4.5\ {\rm \mu m}$ colour cut and measured their auto-power spectra. We then jointly fit models to the three measured spectra to constrain the evolution of the galaxy bias parameter, $b_{\rm g}(z)$. Four bias evolution models were considered: a constant bias model, $b_{\rm g}(z)=b_0$; a constant amplitude model, $b_{\rm g}(z)=b_0/D(z)$; the passive, conserved tracer model of \cite{fry1996}, $b_{\rm g}(z) = 1 + (b_0-1)/D(z)$; and a model linear in redshift, $b_{\rm g}(z)=b_0+b_1z$. 

First, we fit an effective bias model for the mixed radio population without separating the AGN and SFG contributions. Using the Bayesian evidence, the constant bias model is slightly disfavoured, with $\Delta \ln(\mathcal{Z}) = 2.30$ relative to the constant amplitude model, indicating that the data favour a bias that increases with redshift. The passive evolution model is also marginally disfavoured relative to the constant amplitude model, with $\Delta \ln(\mathcal{Z})=1.46$, suggesting the observed evolution may be stronger than expected for a conserved tracer population evolving solely under gravity. The constant amplitude model gives $b_0=1.20^{+0.12}_{-0.14}$ and provides a good fit to the data ($\chi^2_\nu=1.13$). The linear model gives the lowest reduced $\chi^2$ ($\chi^2_\nu=0.95$), but its evidence is comparable to that of the constant amplitude model ($\Delta \ln(\mathcal{Z}) = 0.31$), so the data do not distinguish between these models. The linear model nevertheless prefers a steeper evolution of the effective bias, with $b_1=1.62^{+0.37}_{-0.51}$. This strong evolution may reflect changes in the composition of the radio population with redshift as the relative contributions of AGNs and SFGs evolve and the two populations have different biases.

Using the $1.4\ {\rm GHz}$ radio luminosity functions of \cite{thykkathu2026}, we estimated the fraction of the radio population that are AGNs as a function of redshift. This allowed us to decompose the effective bias into contributions from AGNs and SFGs and model the clustering of these populations separately. Radio-AGNs are found to be significantly more biased tracers of the underlying matter distribution than SFGs, consistent with their association with massive haloes across cosmic time \citep{magliocchetti2017, hale2018, hamlett2026}. Though the evidence does not significantly favour any single model, the constant amplitude evolution is the best-fitting model (smallest reduced $\chi^2$) with $b_{0, {\rm AGN}} = 2.60^{+0.65}_{-0.91}$ and $b_{0, {\rm SFG}} = 0.52^{+0.47}_{-0.34}$. Our inferred AGN bias is somewhat higher than that measured by \cite{hamlett2026} using the same MIGHTEE data. However, their AGN sample was selected using a high luminosity threshold which preferentially selects a larger fraction of HERGs, known to be less strongly clustered than LERGs. We also performed fits in which the SFG bias was constrained using measurements from \cite{hale2025b} as a prior. This tightens the posteriors to $b_{0, {\rm AGN}} = 1.96^{+0.37}_{-0.40}$ and $b_{0, {\rm SFG}} = 0.90^{+0.06}_{-0.06}$.

The main advantage of our approach is that the bias of AGNs and SFGs can be inferred without requiring individual source classifications. Such classifications typically require luminosity cuts or multi-wavelength diagnostics, which can substantially reduce the available sample size. By instead modelling the redshift-dependent population fractions, we are able to use the full radio source catalogue down to low flux densities. In the future, this approach could also be applied to wide-area radio surveys, such as LoTSS \citep{shimwell2026} and EMU \citep{norris2011}, where deep multi-wavelength observations are not available across the entire footprint.

The MIGHTEE fields are well suited to this analysis because they overlap with several multi-wavelength surveys. In particular, \textit{Spitzer} data allowed for the construction of the low- and high-redshift samples, while near-infrared and optical data provided the photometric redshift estimates used to determine the redshift distributions of the radio sources. Galaxy bias measurements depend on accurate knowledge of the redshift distribution of the source population, and future spectroscopic surveys, such as WEAVE-LOFAR \citep{smith2016}, will provide redshifts for large samples of radio sources, allowing these distributions to be determined more accurately.

\section*{Acknowledgements}

JH acknowledges funding from the Science and Technology Facilities Council (STFC) [grant code ST/Y509474/1]. MJJ acknowledges support from the Hintze Family Charitable Foundation through the Oxford Hintze Centre for Astrophysical Surveys. CLH acknowledges support from STFC [ST/Y000951/1]. MJJ and SLJ acknowledge the support of the STFC consolidated grants [ST/S000488/1] and [ST/W000903/1]. MJJ, SLJ, NS and RGV acknowledge funding from a UKRI Frontiers Research Grant [EP/X026639/1]. DA acknowledges support from the Beecroft Trust.

The MeerKAT telescope is operated by the South African Radio Astronomy Observatory, which is a facility of the National Research Foundation, an agency of the Department of Science and Innovation. We acknowledge the use of the ilifu cloud computing facility – www.ilifu.ac.za, a partnership between the University of Cape Town, the University of the Western Cape, Stellenbosch University, Sol Plaatje University and the Cape Peninsula University of Technology. The Ilifu facility is supported by contributions from the Inter-University Institute for Data Intensive Astronomy (IDIA – a partnership between the University of Cape Town, the University of Pretoria and the University of the Western Cape, the Computational Biology division at UCT and the Data Intensive Research Initiative of South Africa (DIRISA). The authors acknowledge the Centre for High Performance Computing (CHPC), South Africa, for providing computational resources to this research project.

We thank the anonymous referee for their useful comments which have helped to improve the manuscript.

\section*{Data Availability}

The MIGHTEE continuum DR1 data used in this work are accessible through \href{https://doi.org/10.48479/7msw-r692}{https://doi.org/10.48479/7msw-r692} with information given in \cite{hale2025}. The cross-matched radio catalogues were released alongside \cite{hale2026}, and the multi-wavelength data catalogues are available by \cite{stylianou2026}. The ACT DR6 lensing maps and simulations used in this work are available at \href{https://portal.nersc.gov/project/act/dr6\_lensing\_v1/}{https://portal.nersc.gov/project/act/dr6\_lensing\_v1/}. The \textit{Spitzer} data catalogues and masks are accessible from the \href{https://irsa.ipac.caltech.edu/cgi-bin/Gator/nph-scan?submit=Select&projshort=SPITZER}{NASA/IPAC Infrared Science Archive}.



\bibliographystyle{mnras}
\bibliography{references}



\appendix

\section{Maps and masks for the low- and high-redshift samples}
\label{sec:maps_and_masks}

\begin{figure*}
	\includegraphics[width=2\columnwidth]{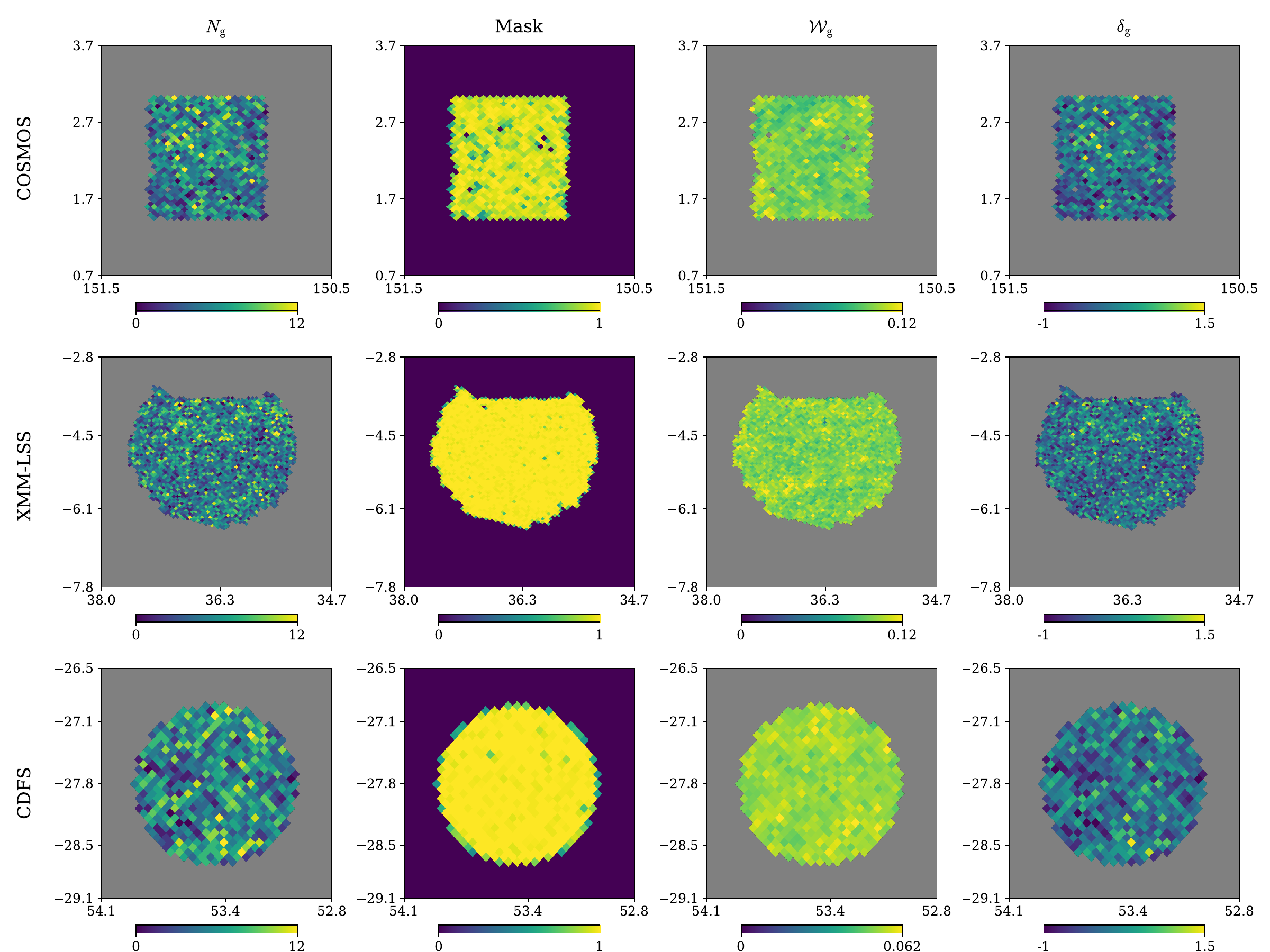}
    \caption{Various maps for the low-redshift radio sample used in this analysis. From left to right, the columns show the radio galaxy number count maps, $N_{\rm g}$, the masks, the galaxy weights maps, $\mathcal{W}_{\rm g}$, and the galaxy overdensity maps, $\delta_{\rm g}$, for each of the MIGHTEE fields (rows). The numbers on the axes are the right ascensions and declinations of the fields in degrees. All colour bars are linear in scale.}
    \label{fig:maps_and_masks_lowz}
\end{figure*}

\begin{figure*}
	\includegraphics[width=2\columnwidth]{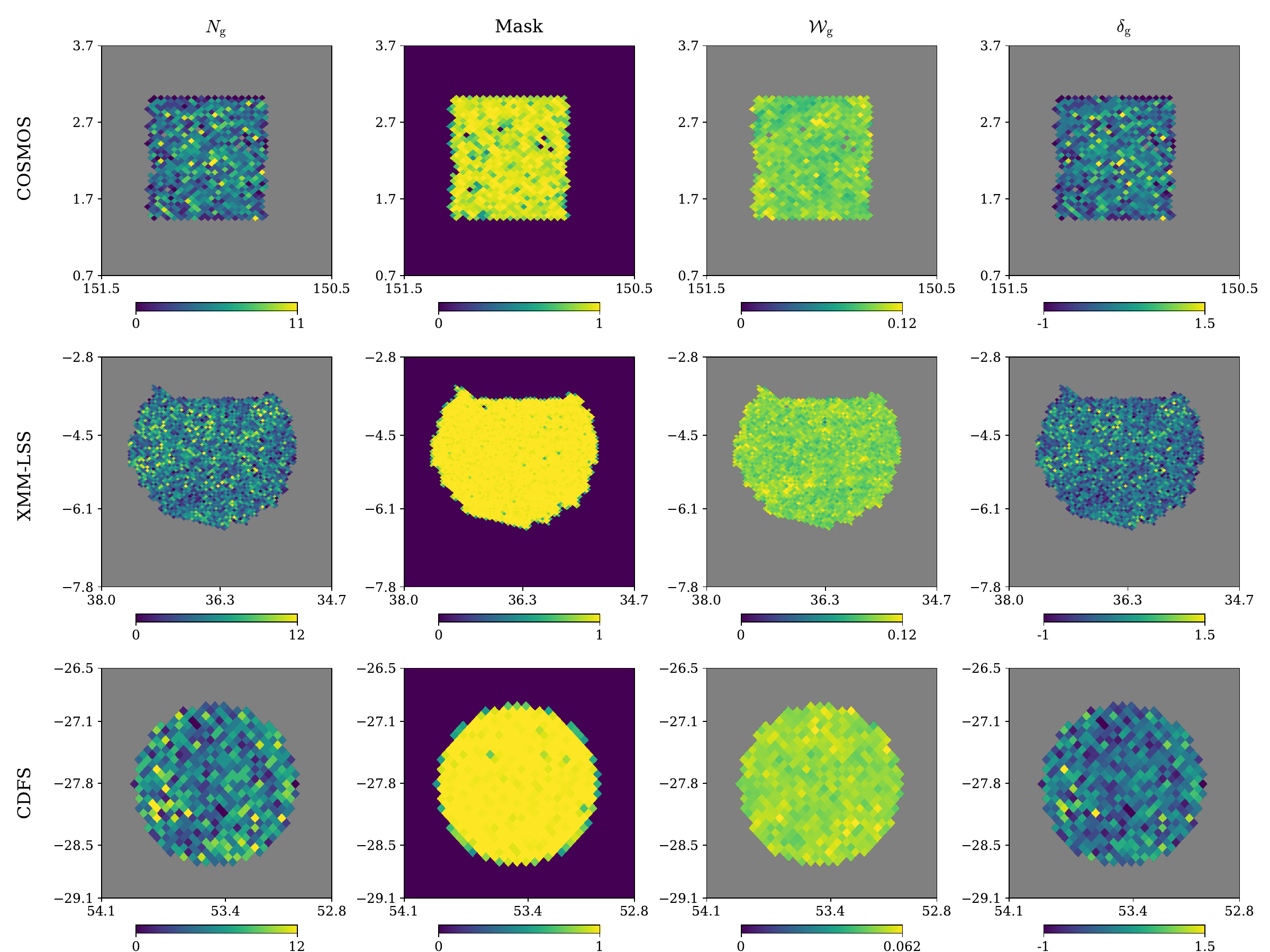}
    \caption{Various maps for the high-redshift radio sample used in this analysis. From left to right, the columns show the radio galaxy number count maps, $N_{\rm g}$, the masks, the galaxy weights maps, $\mathcal{W}_{\rm g}$, and the galaxy overdensity maps, $\delta_{\rm g}$, for each of the MIGHTEE fields (rows). The numbers on the axes are the right ascensions and declinations of the fields in degrees. All colour bars are linear in scale.}
    \label{fig:maps_and_masks_highz}
\end{figure*}

The maps and masks for the low- (Figure~\ref{fig:maps_and_masks_lowz}) and high-redshift (Figure~\ref{fig:maps_and_masks_highz}) radio samples used in this analysis.

\section{Covariance matrix validation}
\label{sec:numerical_covariance}

In Section~\ref{sssec:covariance} we calculated an analytic covariance matrix using \textsc{NaMaster} under the assumption that $\delta_{\rm g}$ and $\kappa$ are Gaussian fields. The limited sky coverage of the radio data ($f_{\rm sky}\approx5\times10^{-4}$) means that the measured bandpowers, particularly at low multipoles, may depart from the Gaussian approximation, with the mask further introducing correlations between modes.

To verify this approximation for the CMB lensing cross-spectrum, we measure the cross-spectrum between the real radio galaxy overdensity map and each of the 400 simulated ACT lensing convergence maps \citep{farren2024}. The components of the covariance matrix are then constructed as
\begin{equation}
    {\rm Cov}_{ij}^{\rm sim} = \frac{1}{N-1} \sum_{r=1}^N \left( \hat{C}_{i}^r - \overline{C}_i \right) \left( \hat{C}_j^r - \overline{C}_j \right), 
\end{equation}
where $\hat{C}_i^r$ is the measured cross-spectrum with the $r$th simulated map in the $i$th multipole bin, and $\overline{C}$ is the average cross-spectrum over all $N$ simulations. Since the simulated convergence maps are independent of the real radio galaxy distribution, no physical correlation is expected, and this ensemble of cross-spectra can therefore be used as an additional null test. The simulated maps include realistic lensing reconstruction noise, and we apply the same radio mask used for the data. The scatter among the resulting cross-spectra therefore provides an empirical estimate of the uncertainties, including the effects of the survey geometry.

\begin{figure}
	\includegraphics[width=\columnwidth]{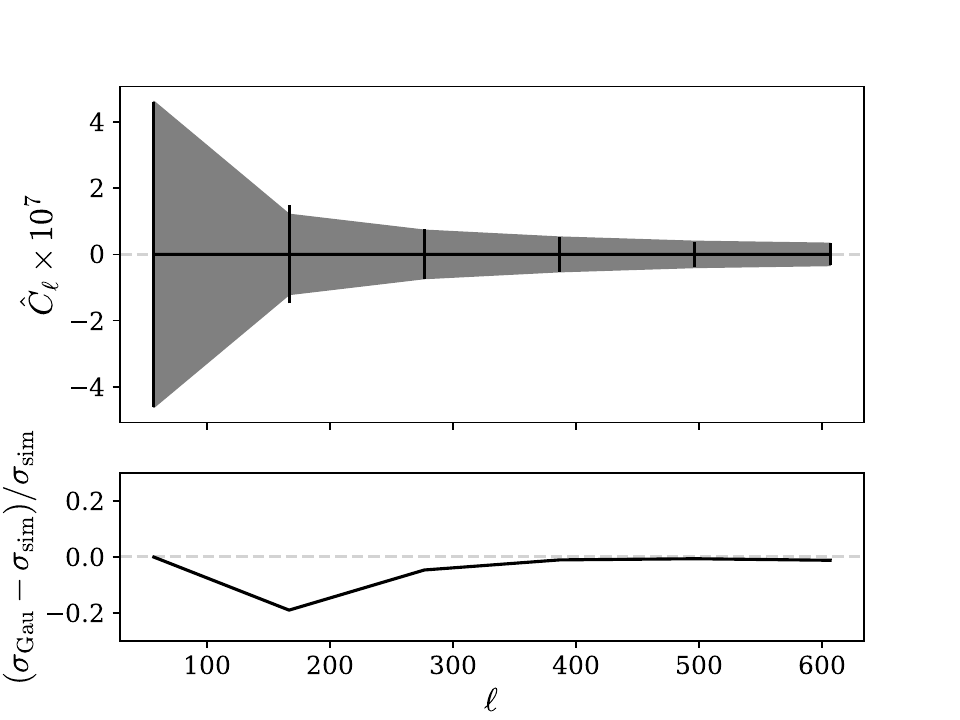}
    \caption{Comparison of the uncertainties on the CMB lensing cross-spectrum, $\hat{C}_\ell$, obtained using the ACT lensing convergence map simulations (black error bars) and the analytic Gaussian covariance matrix (grey shaded region) against multipole, $\ell$. The black line shows the mean power over the 400 simulations. The lower panel shows the fractional difference between the two sets of uncertainties.}
    \label{fig:numerical_covariance}
\end{figure}

Figure~\ref{fig:numerical_covariance} compares the uncertainties estimated from the simulations with those predicted by the analytic covariance. The mean cross-spectrum over the 400 simulations is consistent with zero, confirming that no spurious signal is introduced by our analysis pipeline. The two sets of uncertainties are in good agreement, with differences of less than 5 per cent in all but the second multipole bin where the analytic uncertainty is $\sim20$ per cent smaller than that estimated from the simulations. We repeated the model fitting after inflating the analytic covariance for this bin. The resulting median parameter values and 16th and 84th percentiles of the posteriors change only by a few per cent, and the change in the difference in Bayesian evidence between models would not affect our discussion. We therefore find that the analytic covariance provides a good description of the measurement uncertainties, with only a modest discrepancy in one bin. The off-diagonal elements of the numerical correlation matrix are all $<0.05$, as with the analytic version.
 
\section{Linear bias robustness tests}
\label{sec:lmax_verification}

In Section~\ref{sssec:likelihood}, we determine $\ell_{\rm max}$ using an effective redshift to approximately impose a maximum physical wavenumber of $k_{\rm max}=0.15\ {\rm Mpc}^{-1}$. However, this cut will not be exact for our broad redshift distributions and finite bandpower windows. The redshift distributions extend to $z=0$, so the retained bandpowers can receive contributions from $k>k_{\rm max}$ even when their effective multipoles satisfy the scale cut. For each bandpower, $b$, we therefore quantify the fraction of the theoretical signal arising from small scales as
\begin{equation}
    f_b^{k>k_{\rm max}} = \frac{\sum_\ell W_{b\ell}C_\ell^{k>k_{\rm max}}}{\sum_\ell W_{b\ell}C_\ell},
\end{equation}
where $W_{b\ell}$ is the bandpower window function calculated by \textsc{NaMaster} and $C_\ell^{k>k_{\rm max}}$ is obtained by restricting the Limber integral to the portion where $k(z,\ell)>k_{\rm max}$.

We find substantial high $k$ contribution at the highest retained bandpowers, with $f_b^{k>k_{\rm max}} = 0.43, 0.28$ and $0.38$ for $C_{\ell}^{{\rm full},\kappa}$, $C_{\ell}^{{\rm low}\text{-}z,{\rm low}\text{-}z}$ and $C_{\ell}^{{\rm high}\text{-}z,{\rm high}\text{-}z}$, respectively. At higher multipoles, the fractions exceed 50 per cent. Although these fractions are non-negligible, the presence of $k>k_{\rm max}$ contributions does not necessarily imply a significant shift in the inferred linear bias. It is possible that, given the small sky area we analyse, the impact of the non-linear modes is small compared to our large measurement uncertainties. Rather than imposing a more conservative scale cut, we therefore test whether the residual contribution from $k>k_{\rm max}$ could affect the inferred large-scale bias.

\begin{figure}
	\includegraphics[width=\columnwidth]{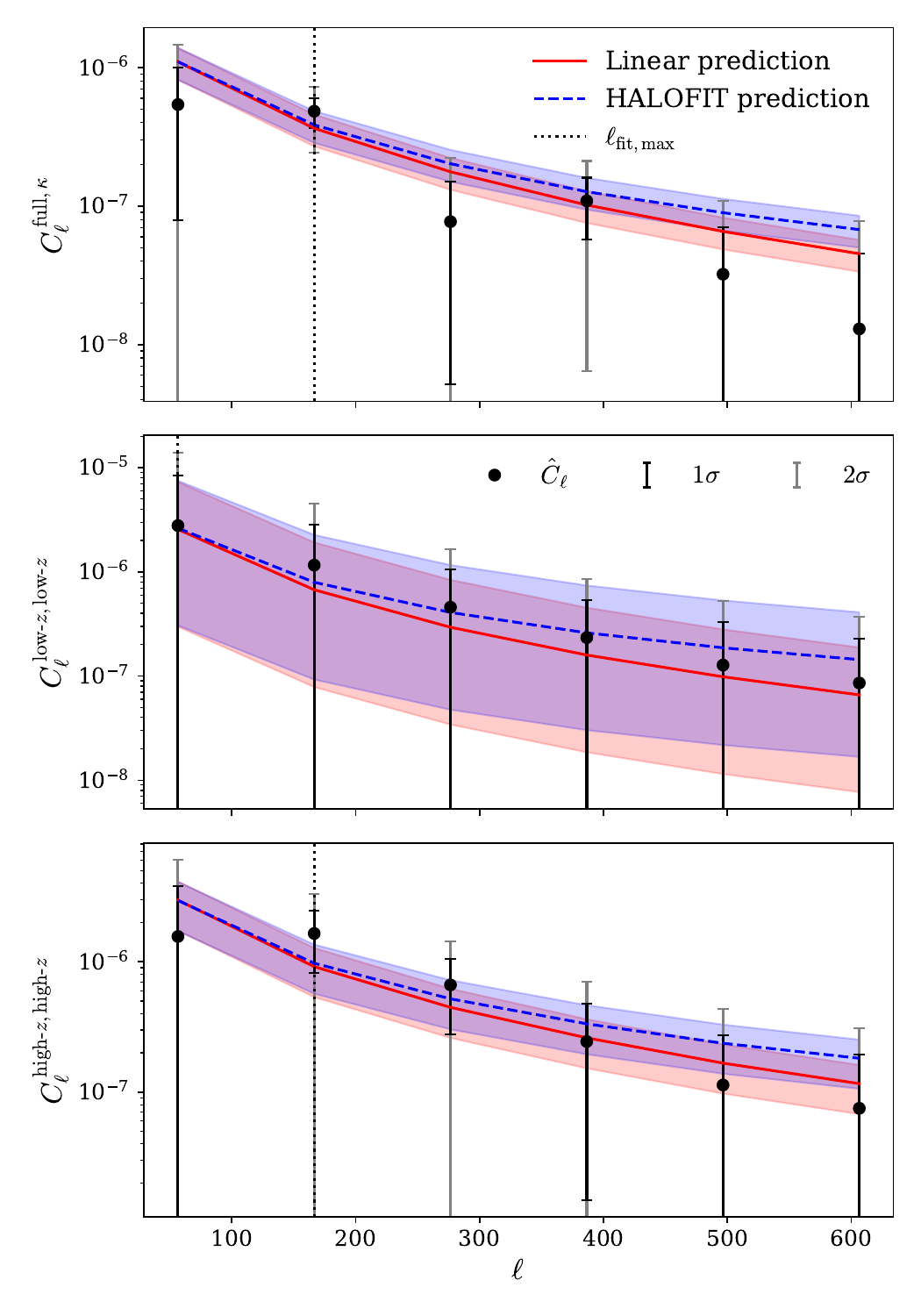}
    \caption{Predicted angular power spectra against multipole, $\ell$, for a linear (solid red lines) and \textsc{HaloFit} (dashed blue lines) matter power spectrum using the bias obtained from fitting to conservative linear scales where the contribution from $k>0.15\ {\rm Mpc}^{-1}$ is less than 10 per cent, $\ell \leq \ell_{\rm fit, max}$ (dotted vertical lines). The data points are the measured angular power spectra, $\hat{C}_\ell$, with the $1\sigma$ (black) and $2\sigma$ (grey) uncertainties. The shaded regions indicate the 16th-84th percentiles of the bias posteriors, and the lines correspond to the median value.}
    \label{fig:linear_halofit_comparison}
\end{figure}

We first assess the sensitivity of our results to non-linear scales by fitting the bias using only conservative scales where the high-$k$ contribution is less than 10 per cent with a linear matter power spectrum. We then use the resulting bias to predict the full multipole range using both the linear and \textsc{HaloFit} matter power spectra. As shown in Figure~\ref{fig:linear_halofit_comparison}, the two predictions differ by less than the $1\sigma$ measurement uncertainties over the full multipole range for all three measured spectra. The \textsc{HaloFit} predictions using this more conservative fit remain consistent with the measured spectra at higher multipoles, with at most a $\sim1\sigma$ difference. This indicates that the higher-multipole measurements are consistent with the bias inferred from the conservative linear-scale analysis, and therefore that including these measurements in our fiducial \textsc{HaloFit} fit does not require a substantially different bias.

As a further test of whether the contribution from $k>k_{\rm max}$ has an impact on the inferred large-scale bias, we allow for scale dependence in the bias. The high-$k$ contribution increases towards higher multipoles, where departures from a scale-independent bias are expected to become increasingly important. If these high-$k$ modes were causing our scale-independent fit to absorb an unmodelled scale dependence, allowing for such dependence should lead to a shift in the inferred large-scale bias, $b_0$. Assuming the bias varies smoothly with wavenumber, the leading-order scale-dependent correction can be written as \citep{maleubre2026}
\begin{equation}
    b(k) = b_0 + b_1 k^2.
\end{equation}
We compare this model with the scale-independent case, $b(k)=b_0$, by progressively increasing $\ell_{\rm max}$ and fitting the model with $b_1$ a free parameter.

\begin{figure}
	\includegraphics[width=\columnwidth]{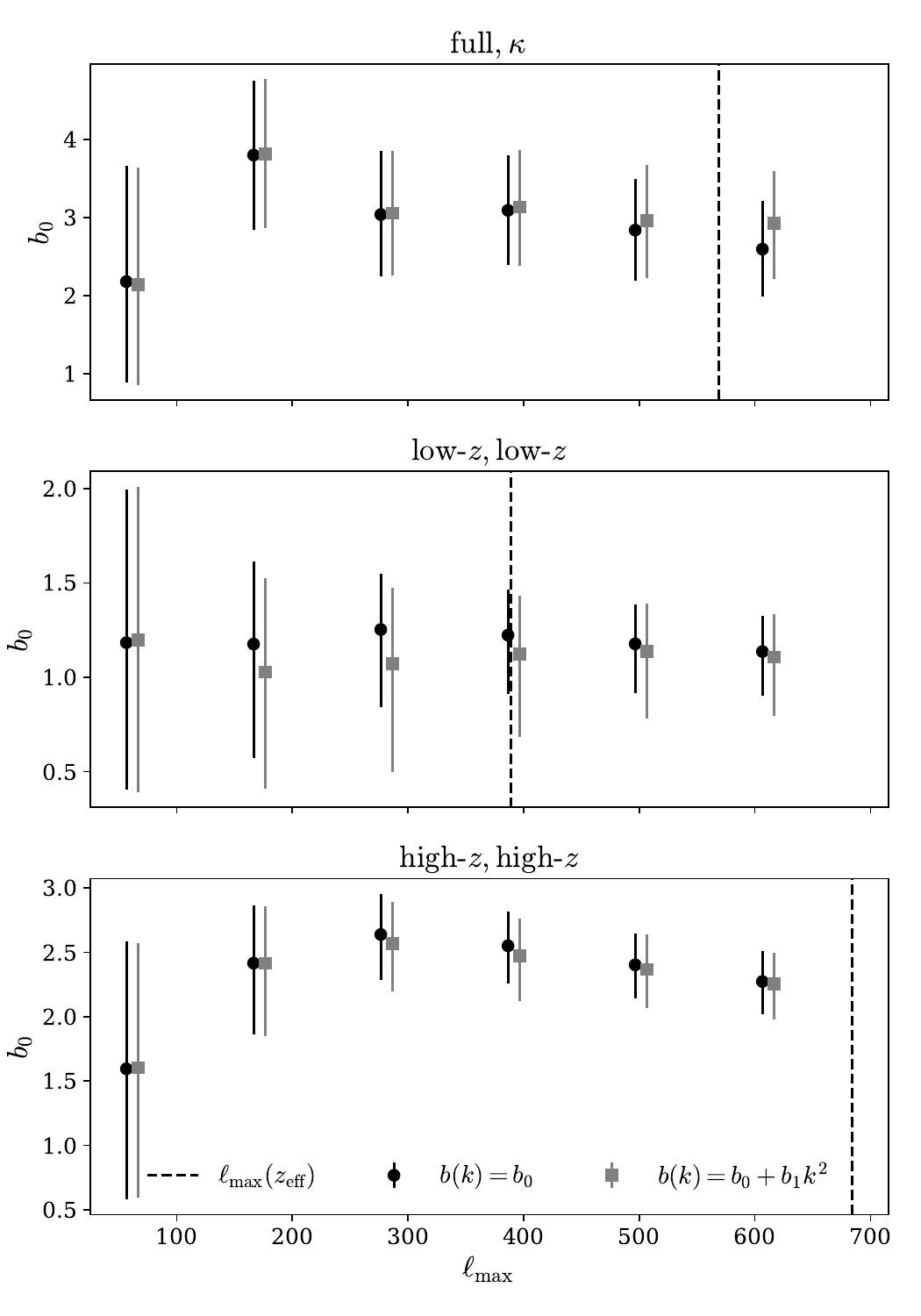}
    \caption{Inferred linear bias, $b_0$, as a function of the maximum multipole included in the fit, $\ell_{\rm max}$, for the three measured power spectra. Points show fits assuming either a scale-independent bias, $b(k)=b_0$ (black circles), or a scale-dependent bias, $b(k)=b_0+b_1k^2$ (grey squares), with the error bars indicating the 16th and 84th percentiles of the posteriors. The vertical dashed lines are the fiducial $\ell_{\rm max}(z_{\rm eff})$ adopted for each spectrum.}
    \label{fig:bias_k2_comparison}
\end{figure}

Figure~\ref{fig:bias_k2_comparison} shows the resulting constraints on $b_0$ as a function of $\ell_{\rm max}$. For all three spectra, $b_0$ from the scale-dependent model remains consistent with that obtained from the scale-independent model over the range of multipoles considered. The inferred $b_0$ is also stable as progressively higher multipoles are included, despite the increasing contribution from modes with $k>k_{\rm max}$. This indicates that the small-scale contributions do not significantly impact our measurements of the linear bias at the precision of our data.

\section{Alternative radio luminosity function}
\label{sec:alternative_luminosity_function}

Table~\ref{tab:agn_sfg_bias_parameters_alternative_luminosity_function} presents the best-fitting AGN and SFG bias model parameters when assuming the alternative luminosity functions of \cite{novak2018} derived from $3\ {\rm GHz}$ VLA-COSMOS data (see Section~\ref{ssec:agn_sfg_bias}).

{\renewcommand{\arraystretch}{1.3}
\begin{table*}
    \centering
    \caption{Best-fitting bias model parameters and auto-spectra shot-noise amplitudes when separating the AGN and SFG contributions to the total effective bias, assuming the alternative luminosity functions of \protect \cite{novak2018}. The corresponding reduced $\chi^2$ values and Bayesian evidence are also reported. The top rows show the results when fitting without using the SFG bias values of \protect \cite{hale2025b} as priors, and the bottom rows show the results when including this prior. The uncertainties correspond to the 16th and 84th percentiles of the posteriors.}
    \begin{tabular}{lcccccccc}
        \hline
        Model & $b_{0, {\rm AGN}}$ & $b_{0, {\rm SFG}}$ & $b_{1, {\rm AGN}}$ & $b_{1, {\rm SFG}}$ & $A^{\rm low}_{\rm sn}$ & $A^{\rm high}_{\rm sn}$ & $\chi^2_\nu$ & $\ln(\mathcal{Z})$ \\
        \hline Without SFG prior \\ \hline
Constant bias & ${4.70}^{+0.83}_{-1.11}$ & ${0.39}^{+0.49}_{-0.29}$ & $-$ & $-$ & ${0.97}^{+0.20}_{-0.20}$ & ${1.00}^{+0.18}_{-0.18}$ & 1.17 & $-11.59\pm0.01$ \\ 
Constant amplitude & ${2.35}^{+0.54}_{-0.74}$ & ${0.50}^{+0.46}_{-0.34}$ & $-$ & $-$ & ${0.98}^{+0.20}_{-0.20}$ & ${0.93}^{+0.19}_{-0.19}$ & 0.95 & $-11.17\pm0.01$ \\ 
Passive & ${2.59}^{+0.42}_{-0.59}$ & ${0.98}^{+0.31}_{-0.17}$ & $-$ & $-$ & ${0.97}^{+0.19}_{-0.20}$ & ${0.98}^{+0.17}_{-0.19}$ & 1.13 & $-12.44\pm0.01$ \\ 
Linear & ${1.33}^{+1.69}_{-0.96}$ & ${0.43}^{+0.49}_{-0.31}$ & ${1.43}^{+1.12}_{-1.13}$ & ${0.80}^{+1.05}_{-0.68}$ & ${0.98}^{+0.20}_{-0.19}$ & ${0.92}^{+0.18}_{-0.19}$ & 1.33 & $-11.93\pm0.01$ \\ 
\hline With SFG prior \\ \hline
Constant bias & ${3.34}^{+0.60}_{-0.71}$ & ${1.10}^{+0.08}_{-0.07}$ & $-$ & $-$ & ${0.97}^{+0.19}_{-0.20}$ & ${1.04}^{+0.18}_{-0.18}$ & 1.44 & $-9.99\pm0.02$ \\ 
Constant amplitude & ${1.83}^{+0.32}_{-0.33}$ & ${0.90}^{+0.06}_{-0.06}$ & $-$ & $-$ & ${0.97}^{+0.20}_{-0.19}$ & ${0.95}^{+0.18}_{-0.18}$ & 1.05 & $-8.71\pm0.02$ \\ 
Passive & ${2.43}^{+0.35}_{-0.37}$ & ${1.10}^{+0.06}_{-0.06}$ & $-$ & $-$ & ${0.97}^{+0.20}_{-0.21}$ & ${0.98}^{+0.18}_{-0.18}$ & 1.19 & $-9.30\pm0.02$ \\ 
Linear & ${1.18}^{+1.23}_{-0.82}$ & ${0.57}^{+0.18}_{-0.18}$ & ${1.39}^{+0.79}_{-1.07}$ & ${1.26}^{+0.39}_{-0.39}$ & ${0.98}^{+0.20}_{-0.20}$ & ${0.93}^{+0.19}_{-0.19}$ & 1.23 & $-8.62\pm0.02$ \\
        \hline
    \end{tabular}
    \label{tab:agn_sfg_bias_parameters_alternative_luminosity_function}
\end{table*}
}

\section{Linear bias model with tied gradients}
\label{sec:linear_tied}

\begin{figure}
	\includegraphics[width=\columnwidth]{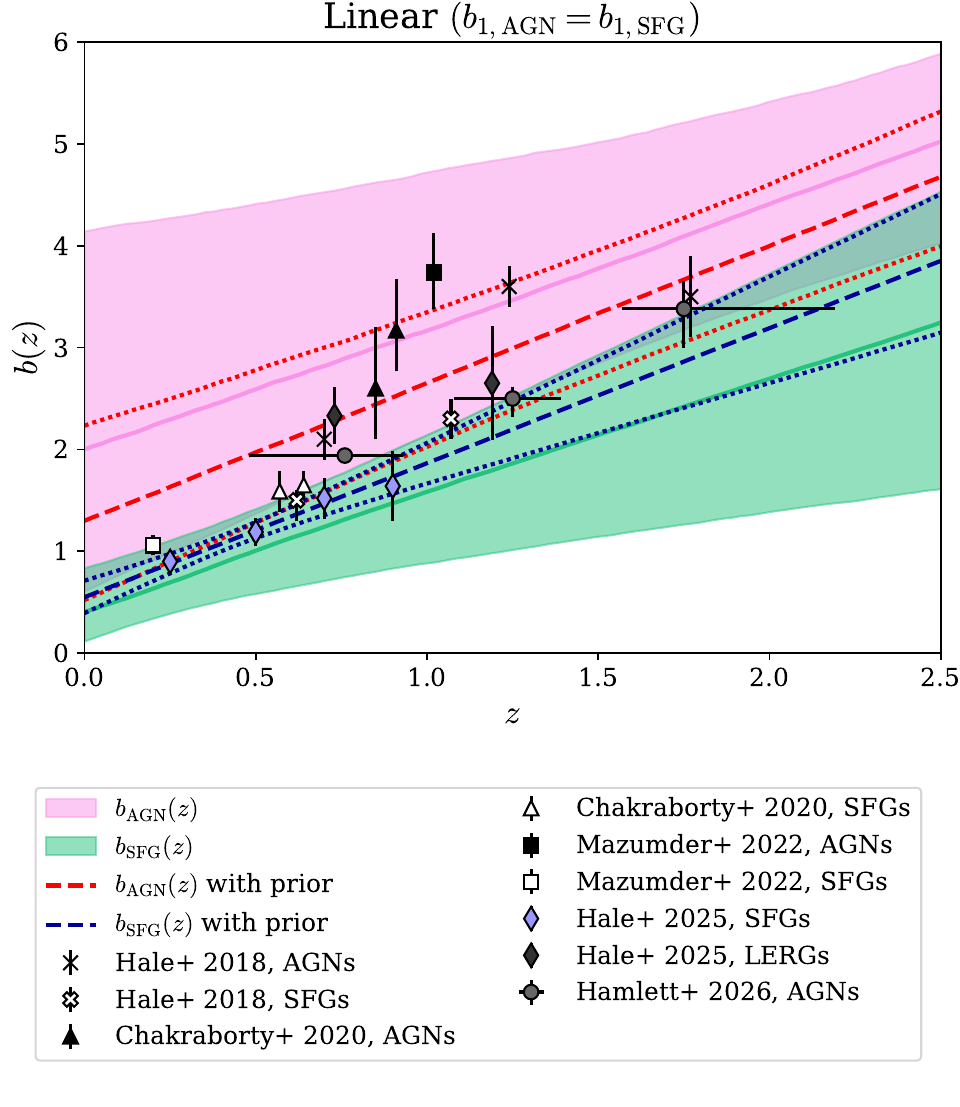}
    \caption{Best-fitting bias evolution for AGNs (pink and red) and SFGs (green and blue) for the linear model with the AGN and SFG gradients tied, such that $b_{1, {\rm AGN}}=b_{1, {\rm SFG}}$. Solid curves show the posterior median predictions, with shaded bands indicating the $1\sigma$ uncertainties, for fits performed without including the SFG bias measurements of \protect \cite{hale2025b} (blue diamonds) as a prior. The dashed and dotted lines are the median posterior predictions and $1\sigma$ uncertainties from fitting with the prior. Also shown are the bias values for LERGs (dark grey diamonds) from \protect \cite{hale2025b} and AGNs (grey circles) from \protect \cite{hamlett2026}, along with the AGN and SFG bias measurements from \protect \cite{hale2018} (crosses), \protect \cite{chakraborty2020} (triangles) and \protect \cite{mazumder2022} (squares). Dark grey and black markers denote AGN bias measurements, while white markers denote SFG bias measurements.}
    \label{fig:agn_sfg_bias_linear_tied}
\end{figure}

We present the results for the linear bias model in which the redshift evolution of the AGN and SFG bias is constrained to have a common gradient, such that $b_{1, {\rm AGN}}=b_{1, {\rm SFG}}$. The resulting best-fitting parameters are listed in Table~\ref{tab:agn_sfg_bias_parameters_linear_tied}, and the corresponding bias evolution is shown in Figure~\ref{fig:agn_sfg_bias_linear_tied}. As discussed in Section~\ref{ssec:agn_sfg_bias}, the fitted common gradient lies between the independently fitted AGN and SFG gradients. Since there is no strong physical motivation for enforcing identical redshift evolution for the AGN and SFG bias, we do not adopt this model in our main analysis. Nevertheless, we include the results here for completeness.

{\renewcommand{\arraystretch}{1.3}
\begin{table*}
    \centering
    \caption{Best-fitting bias model parameters and auto-spectra shot-noise amplitudes when separating the AGN and SFG contributions to the total effective bias and assuming the linear model with the AGN and SFG gradients tied, such that $b_{1, {\rm AGN}}=b_{1, {\rm SFG}}$. The corresponding reduced $\chi^2$ values and Bayesian evidence are also reported. The top row shows the results when fitting without using the SFG bias values of \protect \cite{hale2025b} as priors, and the bottom row shows the results when including this prior. The uncertainties correspond to the 16th and 84th percentiles of the posteriors.}
    \begin{tabular}{lccccccc}
        \hline
        & $b_{0, {\rm AGN}}$ & $b_{0, {\rm SFG}}$ & $b_{1, {\rm AGN}}=b_{1, {\rm SFG}}$ & $A^{\rm low}_{\rm sn}$ & $A^{\rm high}_{\rm sn}$ & $\chi^2_\nu$ & $\ln(\mathcal{Z})$ \\
        \hline
Without SFG prior & ${2.04}^{+1.99}_{-1.45}$ & ${0.40}^{+0.43}_{-0.28}$ & ${1.08}^{+0.52}_{-0.62}$ & ${0.98}^{+0.20}_{-0.19}$ & ${0.94}^{+0.18}_{-0.18}$ & 1.13 & $-13.20\pm0.01$ \\ 
With SFG prior & ${1.34}^{+0.90}_{-0.77}$ & ${0.55}^{+0.16}_{-0.15}$ & ${1.31}^{+0.30}_{-0.33}$ & ${0.98}^{+0.19}_{-0.21}$ & ${0.93}^{+0.18}_{-0.18}$ & 1.10 & $-7.89\pm0.02$ \\ 
        \hline
    \end{tabular}
    \label{tab:agn_sfg_bias_parameters_linear_tied}
\end{table*}
}

\bsp	
\label{lastpage}
\end{document}